\documentclass[nofootinbib,twocolumn,aps,pre,preprintnumbers,floatfix,amsmath,amssymb]{revtex4-1}

\usepackage{graphicx,color,amsmath,bm}
\newcommand{\D}{\mathrm{d}}
\newcommand{\e}{\mathrm{e}}
\newcommand{\half}{\frac{1}{2}}
\newcommand{\vecr}{{\bf r}}
\newcommand{\be}{\begin{equation}}
\newcommand{\ee}{\end{equation}}
\newcommand{\bea}{\begin{eqnarray}}
\newcommand{\eea}{\end{eqnarray}}

\newcommand{\kb}{k_{\mathrm{B}}}
\newcommand{\kbt}{k_{\mathrm{B}}T}

\newcommand{\lb}{l_\mathrm{B}}

\newcommand{\kd}{\kappa_\mathrm{D}}
\begin{document}
%%%%%%%%%%%%%%%%

%\baselineskip=24pt

%%%%%%%%%%%%%%%%%%%%%%%%%%%%%%%%%%%%%%%%%%%%%%%
%\title{A Self-consistent study of Electrolyte Solutions: surface tension at the air/water and oil/water interfaces}
%\title{Surface tension of electrolyte interfaces: a self-consistent study}
%\title{A Self-consistent Theory of Ionic Specific Region in Coulombic Fluids: surface tension at the air/water and oil/water interfaces}

\title{Surface Tension of Electrolyte Interfaces:  Ionic Specificity within a Field-Theory Approach}

\author{Tomer Markovich} \author{David Andelman}
%\email{andelman@post.tau.ac.il}
\affiliation{Raymond and Beverly Sackler School of Physics and Astronomy\\ Tel Aviv
University, Ramat Aviv, Tel Aviv 69978, Israel}

\author{Rudi Podgornik}
\affiliation{Department of Theoretical Physics, J. Stefan Institute,
and \\
Department of Physics, Faculty of Mathematics and Physics\\ University of Ljubljana, 1000 Ljubljana, Slovenia}
%\eads{\mailto{benyaa@post.tau.ac.il}, \mail

\date{accepted to J. Chem. Phys. --- January 2015}

\begin{abstract}
We study the surface tension of ionic solutions at air/water and oil/water interfaces.
By using field-theoretical methods and including a finite
proximal surface-region with ionic-specific interactions.
The  free energy is expanded to first-order in a loop expansion beyond the mean-field result.
We calculate the excess surface tension and obtain analytical predictions that reunite the Onsager-Samaras pioneering result (which does not agree with experimental data), with the ionic specificity of the Hofmeister series.
We derive analytically the surface-tension dependence on the ionic strength, ionic size and ion-surface interaction,
and show consequently that the Onsager-Samaras result is consistent with the one-loop correction beyond the mean-field result.
Our theory fits well a wide range of salt concentrations for different monovalent ions using one fit parameter per electrolyte,
and  reproduces the reverse Hofmeister series for anions at the air/water and oil/water interfaces.
\end{abstract}

%%%%%%%%%%%%%%%%%%%%%%%%%%%%%%%%%%%%%%%%%%%%%%%
\maketitle
%%%%%%%%%%%%%%%%%%%%%%%%%%%%%%%%%%%%%%%%%%%%%%%

%%%%%%%%%%%%%%%%%%%%%%%%%%%%%%%%%%%%%%%
\section{Introduction}
%%%%%%%%%%%%%%%%%%%%%%%%%%%%%%%%%%%%%%%

Surface tension of ionic solutions strongly depends  on their salt composition and, in general, increases with ionic strength for low salt concentrations~\cite{Adamson,Pugh}. Wagner~\cite{Wagner} was the first to connect this finding  with the dielectric discontinuity at the air/water interface, suggesting that dielectric image charge interactions could explain the increase in surface tension. This suggestion was implemented in the pioneering work of Onsager and Samaras (OS), who used the Debye--H\"uckel theory for electrolytes~\cite{Debye1923}.
The OS result presents a universal limiting law for the excess surface tension. It depends on the dielectric mismatch at the interface and on the bulk salt concentration~\cite{onsager_samaras},
and implies that the increase in the surface tension would be independent of the ion type. This simplified observation, however, turned out to be violated in many experimental situations~\cite{Kunz_Book}, and led over the years to numerous investigations of non-electrostatic {\it ion-specific} interactions between ions and surfaces~\cite{Dan2011,Kunz_Book}, and their role in modifying the surface tension of ionic solutions~\cite{Pugh}.

In fact, ion-specific effects date back to the late 19th century, when Hofmeister~\cite{hofmeister}
measured the amount of protein precipitation from solution in presence of various salts, and found a
universal series of ionic activity. The same {\it Hofmeister series} emerges in a large
variety of experiments in chemical and biological systems~\cite{collins1985,ruckenstein2003a,kunz2010}.
Among others, they include forces between mica or silica surfaces~\cite{Sivan2009,Sivan2013,pashely},
osmotic pressure in the presence of (bio)macromolecules~\cite{parsegian1992,parsegian1994}, and more specifically,
measurements of surface tension at the air/water and oil/water interfaces~\cite{air_water_2,air_water_3}.
For simple monovalent salts, the surface tension (in particular at the air/water interface) was experimentally found~\cite{Frumkin1924} to depend strongly on the type of anion, while the dependence on the cation type is much weaker.
This finding is consistent with the fact that anion concentration at the air/water interface exceeds that of cations. Furthermore, for halides~\cite{Kunz_Book}, the lighter ions lead to a larger excess in surface tension in
a sequence that precisely corresponds to the {\it reverse} of the Hofmeister series.

Discrepancies between the OS predictions and the observed ion-specific surface tension motivated numerous attempts to modify the original OS model. Here, we limit ourselves to briefly review some of the more recent  works on surface tension of ionic solutions~\cite{dean2004,dean2003,levin200x,levin200y,Netz2010,Netz2012,Netz2013}, which are directly related to the present study.

Dean and Horgan~\cite{dean2004} calculated the surface tension of ionic solutions to first order in a systematic cumulant
expansion, where the zeroth order is equivalent to the Debye-H\"uckel (DH) linear theory~\cite{Debye1923}.
The specific ion-surface interactions are modeled via an ionic exclusion (Stern) layer of finite thickness. Thus, the interaction of ions with the interface contains only a length scale without any energy scale. This model is solved via a formal field-theoretic representation of the partition function.
The OS result is reproduced exactly, and ion-specific effects are described in terms of the thickness of the salt-exclusion layer.
In yet a separate study by the same authors~\cite{dean2003}, a system of two interacting surfaces including a cation-specific short-range surface interaction was addressed. The consequences of this cation-specific interaction on determining the effective charge, as well as the disjoining pressure as a function of the separation between the surfaces were studied in detail.

In a series of papers Levin and coworkers~\cite{levin200x,levin200y,levin2011} calculated the solvation free-energy
of polarizable ions at air/water and oil/water interfaces.
Their mean-field theory (MFT) modifies the Poisson-Boltzmann (PB) theory by adding an ion-surface interaction potential.
The modified potential is based on several ion-surface interaction terms that are added in an {\it ad hoc} way to the PB equation. These interactions include the image charge interaction, Stern exclusion layer,
ionic cavitation energy and ionic polarizability.
While the additional interaction terms may represent some physical mechanisms for ion-specific interaction with the surface, one cannot, in general, simply add such terms to the MFT potential in a self-consistent way.
These terms, which are sometimes mutually exclusive, are neither completely independent nor can they be obtained from a MFT formulation~\cite{Kunz_Book,hofmeister}.

Computing numerically the surface tension for an homologous series of sodium salts, Levin and co-workers fitted their predictions to the Hofmeister series. They used the hydrated size of the sodium ion as a single fit parameter.
Furthermore, the surface tension at the oil/water interface was fitted for another series of potassium salts~\cite{levin2011}.
In order to apply their theory to the different interface, a second fit parameter was used
to account for the dispersion forces at the oil/water interface.

A different line of reasoning was initiated by Netz and coworkers~\cite{Netz2010,Netz2012,Netz2013},
who calculated the surface tension for both charged and neutral surfaces using a two-scale (atomistic and continuum) modeling approach. Explicit solvent-atomistic molecular-dynamics (MD) simulations furnished non-electrostatic ion-specific potentials of mean force. These interaction potentials were then added to the PB theory that provides the electrostatic part of the potential of mean force.
Within this framework, Netz and coworkers were able to show that the polarity of the surface may reverse the order of the Hofmeister series. The fitted results agree well with experiments performed on hydrophobic and hydrophilic surfaces.

Although many works~\cite{dean2004,levin200x,levin200y,levin2011,Netz2010,Netz2012,Netz2013,attard,diamant1996,ninham2001} tried to generalize the seminal OS theory, it nevertheless remains largely misunderstood what is the accurate theoretical framework of the OS theory.
The OS theory makes use of the (linearized) PB equation
in the presence of a planar dielectric boundary, to obtain the one-dimensional image charge potential of mean force, [see Eq.~(4) in Ref.~\onlinecite{onsager_samaras}]. It should be stated that the latter is {\it not} a solution of the one-dimensional PB equation.
In fact, the PB solution cannot describe any image charge effects on its own. This subtle, yet essential point, gets often irreparably lost when generalizations of the OS theory are attempted based on elaborate decorations of the PB equation itself.

If the OS theory is not simply a redressed version of the PB theory, then what exactly is the relation between them?
While the OS theory cannot be obtained from the mean-field theory, it is deduced from the thermodynamic fluctuations of the instantaneous electric fields around the PB solution~\cite{podgornik1988}.
The free energy is expanded in a loop expansion and only the one-loop correction to the MFT result is retained.

While the detailed formal derivation (as shown below) is complex, we believe that its physical basis is quite simple and straightforward. The OS result does not generalize the PB equation to include image charge effects at an interface between two dielectric media, but rather it solves the problem on a higher level of approximation by going beyond the MFT level. We consider this conceptual clarification to have large importance on generalizations of the OS theory itself.

In this paper we introduce two important modifications, relevant to the calculations of the surface tension of electrolytes. First, we demonstrate that the OS contribution is effectively fluctuational in nature, and follows from the one-loop expansion of the Coulomb partition function around the mean-field solution.
Second, we propose a phenomenological approach that consistently describes ion-interface interactions in the form of a coupling term in the free energy.
This new formulation allows us to obtain a simple analytical theory that reunites the OS pioneering result, which does not agree with experimental data, with the ionic specificity of the Hofmeister series.

We take ionic specificity into account through the ionic size and an ion-surface interaction.
Each ionic species is characterized by a phenomenological adhesivity parameter~\cite{Davies,diamant1996,dean2003}.
Specifically, short-range non-electrostatic effects such as the ion chemical nature, size and polarizability, as well as the preferential ion-solvent interaction~\cite{Dan2011,US1,US2,Onuki_curr_op,delacruz2012}, are introduced by adding one phenomenological parameter to the free energy.
This allows us to obtain a modified PB mean-field theory and to evaluate the contribution of fluctuations (beyond mean-field) to the surface tension.
The latter includes the dielectric image charge effects (OS), as well as the couplings between image charge effects and surface-specific interactions.
Our analytical surface tension prediction fits well a variety of interfacial tension data at the air/water and oil/water interfaces.
Using one fit parameter per electrolyte it reproduces the reversed Hofmeister series for several types of monovalent anions.

%%%%%%%% Outlook  %%%%%%%%
The outline of this paper is as follows.
In Sec. II, we present the model and a general derivation of the grand-potential to the one-loop order.
In Sec. III we find the  free energy and treat the spurious divergencies of our model,
while in Sec. IV we derive an analytical expression for the surface tension.
Finally, a comparison of our results with experimental data (Sec. V) and some concluding remarks (Sec. VI) are presented.

%%%%%%%%%%%%%%%%%%%%%%%%%%%%%%%%%%%%%%%
\section{The Model}
%%%%%%%%%%%%%%%%%%%%%%%%%%%%%%%%%%%%%%%

We consider an ionic solution that contains monovalent symmetric ($1$:$1$) salt of charge $\pm e$ and of bulk concentration $n_b$.
The aqueous phase (water) volume $V=AL$ has a cross-section $A$ and an arbitrary large length, $L\to \infty$.
The surface between the aqueous and air phases is chosen at $z=0$.
The two phases are taken as two continuum media with uniform dielectric
constant $\varepsilon_w$ and $\varepsilon_a$, respectively, such that:
\begin{equation}
\label{m1}
\varepsilon({\bf r}) =
\left\{
    \begin{array}{rl}
        \varepsilon_a & ~~~~ z < 0\\
        \varepsilon_w & ~~~~ z \geq 0
    \end{array}
\right. \, .
\end{equation}
Note that we can equally model the water/oil interface. By assuming no ions penetrate the oil phase, we model the oil/water system by taking $\varepsilon_a$ as the dielectric constant of the oil phase.

The ion self-energy will not be considered explicitly because it is well known that this self-energy
is extremely large ($\sim 100\kbt$) in the air phase. It will only be taken into account implicitly by confining the ions into the water phase.
We also assume a proximal region inside the water phase where there are ion-specific interactions of the ions with the interface. The width of this region is denoted by $d$,
and the interactions are modeled by a potential $V_{\pm}(z)$ for anions and cations, respectively.
We also assume that these interactions depend only on the $z$ coordinate, which
is a reasonable assumption if the surface is rather uniform and flat.

%%%%%%%%%%%%%%%%% fig 1%%%%%%%%%%%%%%%%%%%%
%Fig1
\begin{figure}[h!]
\includegraphics[scale=0.6]{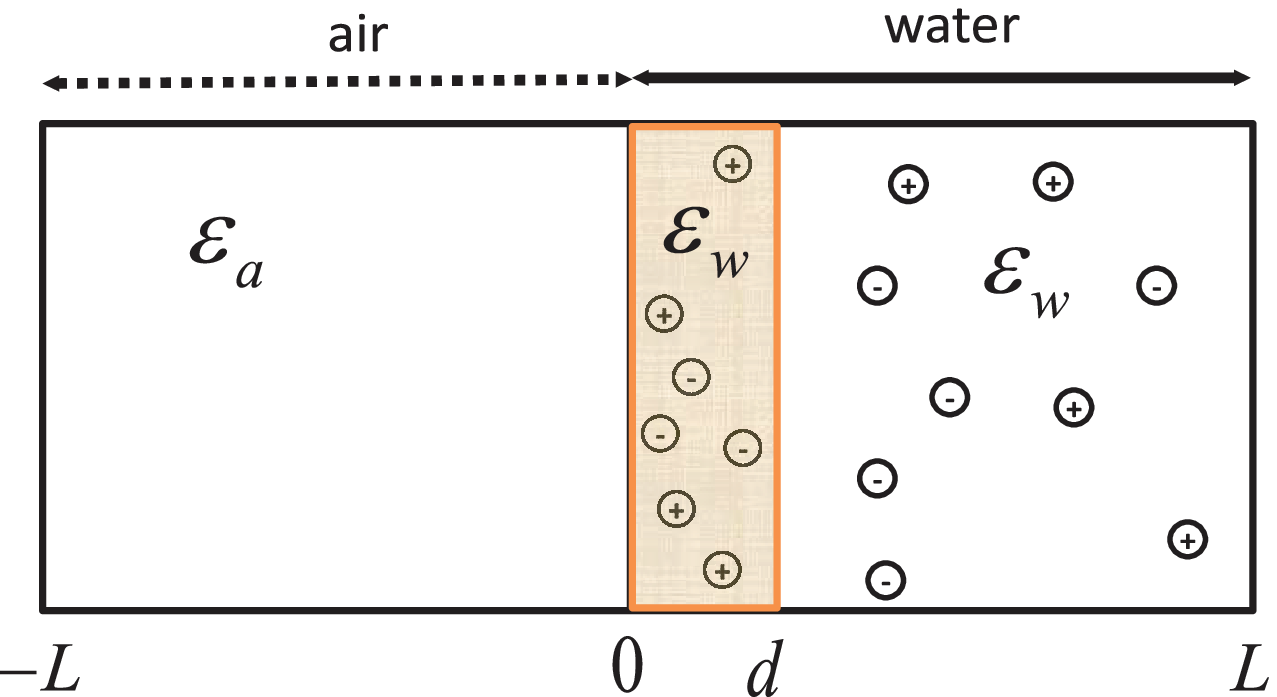}
\caption{\textsf{Schematic image of the system. The water and air phases have the same length $L$, where  $L\rightarrow\infty$. In the water phase, there is a proximal region, $0<z<d$, where the anions and cations interact with a surface interaction, modeled by an external potential $V_{\pm}(z)$.
}}\label{fig1}
\end{figure}
%%%%%%%%%%%%%%%%%%%%%%%%%%%%%%%%%%%%%%%%%%%
%

The model Hamiltonian is:
\begin{eqnarray}
\label{m2}
\nonumber &H =& \half\sum_{i\ne j}q_i q_j u({\bf r}_i,{\bf r}_j)
+ \!\!\!\!\sum_{i\, \epsilon\, {\rm anions}} \!\! V_{-}(z_i) \left[ \theta(z_i) - \theta(z_i - d)\right] \\
& &+ \!\!\!\!\sum_{i\, \epsilon\, {\rm cations}} \!\! V_{+}(z_i) \left[ \theta(z_i) - \theta(z_i - d)\right]  \, ,
\end{eqnarray}
where $q_i=\pm e$ is the charge of monovalent anions and cations, respectively, and we use the Heaviside function,
\begin{equation}
\label{m2a}
\theta(z) =
\left\{
    \begin{array}{rl}
        0 & ~~~~ z < 0\\
        1 & ~~~~ z \geq 0
    \end{array}
\right. \, .
\end{equation}
The first term is the usual Coulombic interaction which satisfy
$\nabla\cdot[\varepsilon({\bf r})\nabla u({\bf r,r^\prime})] = -4\pi\delta({\bf r-r'})$,
where the diverging self-energy of point-like ions is subtracted.
The second and third terms are the ion-surface specific interaction for anions and cations, respectively.

The thermodynamical grand-partition function can be written as
\begin{eqnarray}
\label{m3}
\nonumber& \Xi = &\sum_{N_-=0}^{\infty} \sum_{N_+=0}^{\infty} \frac{\lambda_-^{N_-}}{N_-!} \frac{\lambda_+^{N_+}}{N_+!} \int\prod_{i=1}^{N_-}\D^3r_i \prod_{j=1}^{N_+}\D^3r_j \\
\nonumber &\times& \exp\left(-\frac{\beta}{2}\sum_{i,j}q_i q_j u({\bf r}_i,{\bf r}_j) \right. \\
\nonumber& & \left. -\beta\sum_{i} V_{-}(z_i) \left[ \theta(z_i) - \theta(z_i - d)\right] \right. \\
& &\left. -\beta\sum_{j} V_{+}(z_j) \left[ \theta(z_j) - \theta(z_j - d)\right] \right) \, ,
\end{eqnarray}
where $\beta = 1/\kbt$, $\kb$ is the Boltzmann constant, $T$ is the temperature
and $N_{\pm}$ are, respectively, the number of cations and anions. Note that the sum is over all $i$ and $j$.
The fugacities $\lambda_{\pm}$ are defined via the (intrinsic) chemical potentials, $\mu_\pm$, as
\begin{eqnarray}
\label{m7}
\lambda_{\pm} &=& a^{-3}\exp[{\beta\mu_{\pm}^{\rm tot}}] \, ,\nonumber\\
\mu_{\pm}^{\rm tot}&=& \mu_{\pm}({\bf r}) + \half e^2u_{\rm self}({\bf r,r})\, , \quad
\end{eqnarray}
where the (diverging) self-energy, $u_{\rm self}({\bf r,r})$, of the 1:1 monovalent ions is subtracted,
and $\mu_{\pm}^{\rm tot}$
is the total chemical potential.
The length scale $a$ is a microscopic length corresponding to the ionic size
and is assumed to be equal for anions and cations.

We can then write the charge density operator, $\hat\rho$, as
\begin{eqnarray}
\label{m4}
\hat{\rho}({\bf r}) = \sum_j q_j \, \delta({\bf r}-{\bf r}_j) \, ,
\end{eqnarray}
%
%and $q_j = \pm e$ is the charge of monovalent anions and cations respectively.
and in order to proceed we introduce the functional Dirac delta function,
\begin{eqnarray}
\label{m5}
\delta\left[\rho({\bf r}) \!-\! \hat{\rho}({\bf r})\right]
&=& \left(\frac{\beta}{2\pi}\right)^N \int {\cal D} \phi({\bf r})\\
\nonumber& &\times \exp{\left[i\beta\int\D^3r\,\phi({\bf r})\left(\rho({\bf r}) - \hat{\rho}({\bf r})\right)\right]} \, ,
\end{eqnarray}
with $\phi$ being an auxiliary field and $N=N_+ + N_-$.
The functional integral representation, is a functional integral over all values of $\left\{\phi({\bf r})\right\}$ at all space points ${\bf r}$. It can be thought of as the continuum limit of multiple integrals over values of $\phi({\bf r})$ at different points in space, $\int\prod_{i=1}^{N} \D\phi({\bf r}_i) \to \int {\cal D}\phi({\bf r})$, for $N\to\infty$.
Rewriting the grand-partition function using Eqs.~(\ref{m4})-(\ref{m5}) and the Hubbard-Stratonovich transformation~\cite{podgornik1988} yields:
\begin{eqnarray}
\label{m6}
& &\Xi = \frac{\left(2\pi\right)^{-N/2}}{\sqrt{\det[\beta^{-1} u({\bf r,r'})]}} \\
\nonumber& & \times \int {\cal D}\phi \, \exp\Bigg[-\frac{\beta}{2}\int\D^3r\D^3r'\phi({\bf r})u^{-1}({\bf r,r'})\phi({\bf r'}) \Bigg. \\
\nonumber& & + \Bigg. \int\D^2r \int_{0}^{d}\D z \left( \lambda_+\e^{-i\beta e\phi({\bf r}) - \beta V_{+}(z) } + \lambda_-\e^{i\beta e\phi({\bf r}) - \beta V_{-}(z) }\right) \Bigg. \\
\nonumber& &\Bigg. + \int\D^2r  \int_{d}^{L}\D z \left( \lambda_+\e^{-i\beta e\phi({\bf r}) } + \lambda_-\e^{i\beta e\phi({\bf r})} \right) \Bigg]  \, .
\end{eqnarray}

The grand-partition function can be written in the form:
\begin{eqnarray}
\label{m8}
\Xi \equiv \frac{\left(2\pi\right)^{-N/2}}{\sqrt{\det[\beta^{-1} u({\bf r,r'})]}}\int D\phi\,\e^{-S\left[\phi({\bf r})\right]} \, ,
\end{eqnarray}
where $S\left[\phi({\bf r})\right]$ plays the role of a field action,
\begin{eqnarray}
\label{m9}
& &S\left[\phi({\bf r})\right] = \int\D^3r\,\frac{\beta\varepsilon({\bf r})}{8\pi}[\nabla \phi({\bf r})]^2  \\
\nonumber& & - \int\D^2r \bigg[ \lambda \int_{0}^{d}\D z  \left( \e^{-i\beta e\phi({\bf r}) - \beta V_{+}(z) } + \e^{i\beta e\phi({\bf r}) - \beta V_{-}(z) }\right)  \bigg.  \\
\nonumber& &\Bigg. \bigg. +\, 2\lambda \int_{d}^{L}\D z \cos\left[\beta e \phi({\bf r}) \right]  \bigg]  \, ,
\end{eqnarray}
with $\varepsilon({\bf r})$ is the dielectric function defined in Eq.~(\ref{m1}). In the above equation we have also used the inverse Coulomb potential $u^{-1}({\bf r},{\bf r}')=-\frac{1}{4\pi} \nabla\cdot[\varepsilon({\bf r})\nabla\delta({\bf r}-{\bf r}')]$ that obeys
$\int\D^3 r'' u(\vecr,\vecr'')u^{-1}(\vecr'',\vecr') = \delta(\vecr-\vecr')$. The electro-neutrality $e(\lambda_{+}-\lambda_{-})=0$  imposes $\lambda_+=\lambda_-\equiv \lambda$.

For slowly varying potentials, $V_{\pm}$ within the proximal region, one can write,
\begin{eqnarray}
\label{m10}
\int_0^d \D z ~f(z) \, e^{- \beta V_{\pm}(z)} \simeq \left < \e^{- \beta V_{\pm}(z)} \right>_z \int_0^d \D z ~f(z) \, ,
\end{eqnarray}
where  $\langle O \rangle_{z}=\frac{1}{d}\int_0^d\D z\, O(z)$  denotes the spatial average in the $[0,d]$ interval.
Indeed, if $V_{\pm}$ vary slowly in the $0<z<d$ region, we can write the field action using Eq.~(\ref{m10}) as,
\begin{eqnarray}
\label{m12}
& &S\left[\phi({\bf r})\right] = \int\D^3r\, \frac{\beta\varepsilon({\bf r})}{8\pi}[\nabla \phi({\bf r})]^2 \\
\nonumber& & - \int\D^2r \left( \lambda \int_{0}^{d}\D z  \left( \e^{-i\beta e\phi({\bf r}) - \beta \alpha_{+} } + \e^{i\beta e\phi({\bf r}) - \beta \alpha_{-} }\right)   \right. \\
\nonumber& & \left. +\, 2\lambda \int_{d}^{L}\D z \cos\left[\beta e \phi({\bf r})  \right] \right)  \, ,
\end{eqnarray}
where $\e^{- \beta \alpha_{\pm}} \equiv \left< \e^{- \beta V_{\pm}(z)} \right>_{z}$, or to first-order in a cumulant expansion,
$\alpha_\pm=\langle V_\pm\rangle_z$.

The above equation represents the full grand-partition function of our electrolyte system as represented by a field theory. Because of the existence of specific ion-surface interactions, it contains additional length scales besides the electrostatic ones. These additional length scales depend on $\alpha_{\pm}$ and do not allow to introduce a single electrostatic coupling parameter, which leads to the weak- and strong-coupling limits~\cite{podgornik_review}. Nevertheless, we can still introduce the weak-coupling limit as it corresponds to a saddle-point configuration with Gaussian fluctuations. This is consistent not only with weak electrostatic interactions but also with weak specific surface-ion interactions.
Up to first-order in a loop expansion, the field action, Eq.~(\ref{m9}), can be written as:
\begin{eqnarray}
\label{m13}
\nonumber &S\left[\phi({\bf r})\right] &\simeq S\left[\phi_0{\bf r})\right] \\
\nonumber& & +\, \frac{1}{2}\int\left.\frac{\delta^2S\left[\phi({\bf r})\right]}{\delta\phi({\bf r})\delta\phi({\bf r}')}\right|_0 \delta\phi({\bf r})\delta\phi({\bf r}')\,\D^3r\D^3r' \,.\\
\end{eqnarray}

It is useful to define the Hessian of $S$ as:
\begin{equation}
H_2({\bf r,r'}) = \frac{\delta^2S}{\delta\phi({\bf r}')\delta\phi({\bf r}'')}\Big|_{_0}\,.
\end{equation}
The subscript `$0$' stands for the value of the field action at its stationary point defined as $\left.\delta S / \delta\phi({\bf r})\right|_0 = 0$, and $\phi_0({\bf r})$ is the value of $\phi({\bf r})$ at this point.
The functional integral in Eq.~(\ref{m10}) is Gaussian and can be evaluated explicitly giving the grand potential, $\Omega = -\kbt\ln\Xi$, in the form
\begin{eqnarray}
\label{m14}
\nonumber & \Omega & \simeq \Omega_0+\Omega_1 \\
& &= \kbt S_0 + \half\kbt\, {\rm Tr}\ln\left(H_2({\bf r,r'}) \right)\, ,
\end{eqnarray}
where we have dropped irrelevant constant terms and used the matrix identity, $\ln\det\left(A\right) = {\rm Tr}\ln \left(A\right)$.
%
%

%%%%%%%%%%%%%%%%%%%%%%%%%%%%%%%%%%%%%%%%%%%%%%%%%%%%
\subsection{Mean Field Theory}
%%%%%%%%%%%%%%%%%%%%%%%%%%%%%%%%%%%%%%%%%%%%%%%%%%%%

The MFT equation corresponds to the saddle-point of $S$, and can be written for the mean-field electrostatic potential $\psi({\bf r})$ by identifying $\psi({\bf r}) = i\phi_{0}({\bf r})$.
The stationary point of Eq.~(\ref{m12}) then implies
\begin{eqnarray}
\label{mf1}
\nonumber &\nabla^2\psi_1 = 0 \quad &z<0 \\ \nonumber \\
\nonumber &\nabla^2\psi_2 =
\frac{4\pi en_b}{\varepsilon_w}\left( \e^{-\beta\alpha_{-}}\e^{\beta e\psi_2} -
\e^{-\beta\alpha_{+}}\e^{-\beta e\psi_2} \right) \quad &0\leq z \leq d \\ \nonumber \\
\nonumber &\nabla^2\psi_3= \frac{8\pi en_b}{\varepsilon_w}\sinh \left( \beta e\psi_3 \right) \quad  &z>d \\
\end{eqnarray}
and is equal to the standard PB equation in the $z>d$ region.
The three spatial regions are denoted by `1' for $z<0$ (air), `2' for $0\leq z \leq d$ (proximal layer) and `3' for $z>d$ (distal region).
Note that on the mean-field level, the fugacity $\lambda$ can be replaced by the bulk
salt concentration, $n_b$ (see Appendix~C).

Because of the $(x,y)$ in-plane translation symmetry, the MFT potential varies only in the perpendicular $z$-direction, $\psi({\bf r}) = \psi(z)$. We simplify the treatment by linearizing the MFT equations (the DH limit), and obtain
\begin{eqnarray}
\label{mf2}
\nonumber& &\psi_1'' = 0 \, ,\\\nonumber \\
\nonumber& &\psi_2'' = \frac{1}{\beta e}\frac{\kd^2}{2}\left( \e^{-\beta\alpha_{-}} - \e^{-\beta\alpha_{+}} \right) + \xi^2\psi_2 \, , \\\nonumber \\
& &\psi_3'' = \kd^2\psi_3 \, ,
\end{eqnarray}
with
\begin{equation}
\label{xidef}
\xi^2 \equiv \frac{\kd^2}{2}\left(\e^{-\beta\alpha_{-}} + \e^{-\beta\alpha_{+}}\right),
\end{equation}
and the inverse Debye length defined as
\begin{equation}
\label{mf2b}
\kd = \sqrt{8\pi\beta e^2 n_b /\varepsilon_w} \, .
\end{equation}
The linearization is valid for a surface potential that is rather small, $\left|\beta e\psi\right|\ll1$,
and corresponds to $\alpha_{-} \simeq \alpha_{+}$.

Using the continuity of the electrostatic potential and its derivative at $z=0$ and $z=d$, with vanishing electrostatic field in the bulk ($z \to L$) and in the air phase $(z<0)$ we obtain the linear MFT solution:
\begin{eqnarray}
\label{mf3}
\nonumber& &\psi_1 = \frac{ \kd\left(\cosh\xi d - 1\right) + \xi\sinh \xi d}{\kd\cosh\xi d + \xi\sinh \xi d} \chi \, ,\\ \nonumber \\
\nonumber& &\psi_2 = \frac{ \kd\left(\cosh\xi d - \cosh\xi z\right) + \xi\sinh \xi d}{\kd\cosh\xi d + \xi\sinh \xi d} \chi \, , \\ \nonumber \\
& &\psi_3 = \frac{\xi\sinh \xi d}{\kd\cosh\xi d + \xi\sinh \xi d} \e^{-\kd (z-d)} \chi \, ,
\end{eqnarray}
with the parameter $\chi$ defined as
\begin{eqnarray}
\label{mf3a}
\beta e\chi \equiv \frac{\e^{-\beta\alpha_{+}} - \e^{-\beta\alpha_{-}}} { \e^{-\beta\alpha_{-}} + \e^{-\beta\alpha_{+}} }\, .
\end{eqnarray}
Notice that for $\alpha_{\pm}=0$, the parameter $\chi=0$, the surface at $z=d$ is a phantom surface, and $\psi(z)$ vanishes everywhere.
Another limit where $\psi=0$ everywhere is obtained for $d\to 0$ (no proximal region).

Keeping only linear terms in $d$, the electrostatic potential is,
\begin{eqnarray}
\label{mf5}
\nonumber& &\beta e\psi_1 = \kd d \left( \e^{-\beta\alpha_{+}} - \e^{-\beta\alpha_{-}} \right) \, , \\
\nonumber& &\psi_2 = \psi_1 \, ,\\
& &\psi_3 = \psi_1\e^{-\kd z} \, .
\end{eqnarray}
By taking $\alpha_{+} = 0$, while keeping $\alpha_{-}\ne 0$ in the above equation,
the result of Ref.~\onlinecite{EPL} is recovered for a single type of adsorbing ion
subjected to a surface interaction.
This is equivalent to the limit of a proximal layer  of zero width.

The approximation for the field action is then
\begin{eqnarray}
\label{mf4}
\nonumber&\Omega_0 & = -\int\D^3r\, \frac{\beta\varepsilon({\bf r})}{8\pi}[\nabla \psi({\bf r})]^2  \\
\nonumber& & - \kbt n_b \int\D^2r \int_{0}^{d}\D z  \left( \e^{-\beta e\psi - \beta \alpha_{+} } + \e^{\beta e\psi - \beta \alpha_{-} }\right)   \\
& & - 2\kbt n_b \int\D^2r   \int_{d}^{L}\D z \cosh\left(\beta e \psi \right)    \, ,
\end{eqnarray}
where the MFT potential is obtained from Eq.~(\ref{mf3}) and $\psi(\vecr) = i\phi_0(\vecr)$.

%%%%%%%%%%%%%%%%%%%%%%%%%%%%%%%%%%%%%%%%%%%%%%%%%%%%
\subsection{One-Loop Correction}
%%%%%%%%%%%%%%%%%%%%%%%%%%%%%%%%%%%%%%%%%%%%%%%%%%%%

In order to obtain the one-loop correction, one should first calculate the trace of the logarithm of the Hessian (see Eq.~(\ref{m14})).
To do so we start by considering the eigenvalue equation of the Hessian:

\begin{equation}
\label{ol1}
\int \D^3r'\,H_2({\bf r,r'})u_{\nu}({\bf r'})=\nu u_{\nu}({\bf r}) \, .
\end{equation}
Because of the planar symmetry of our system, ${\bf r}=(\bm{\rho},z)$, where $\bm{\rho}$ is the inplane vector, it is convenient to write down the eigenvalue problem in the transverse Fourier space, where ${\bf k}$ is coupled to $\bm{\rho}$ and  $U_{\nu}({\bf k},z)$ is the transverse Fourier transform of $u_\nu(\bm{\rho},z)$
\begin{equation}
\label{ol2}
u_\nu(\bm{\rho},z)=\frac{A}{4\pi^2}\int \D^2{\bf k} \,U_\nu({\bf k},z) \e^{i{\bf k} \cdot \bm{\rho} } \, .
\end{equation}
In Fourier space, the eigenvalue equation is
\begin{eqnarray}
\label{ol3}
& &\Bigg( -  \varepsilon_w\kd^2\cosh(\beta e\psi) \left[ \theta(z-d) - \theta(z-L) \right] \Bigg. \\
\nonumber& &\Bigg. - \half \varepsilon_w\kd^2\left(  \e^{-\beta\alpha_{-}}\e^{\beta e\psi} + \e^{-\beta\alpha_{+}}\e^{-\beta e\psi}  \right)\left[ \theta(z) - \theta(z-d) \right] \Bigg. \\
\nonumber& &\Bigg.  + \partial_z\varepsilon(z)\partial_z -  \varepsilon(z)k^2   \Bigg)U_{\nu}({\bf k},z)=\nu U_\nu({\bf k},z)  \, ,
\end{eqnarray}
where $\varepsilon(\vecr) = \varepsilon(z)$ as in Eq.~(\ref{m1}),
and the corresponding boundary condition at $z=0$ is
\begin{eqnarray}
\label{ol4}
& & \varepsilon_w \partial_z U^{(2)}_{\nu}\bigg|_{0^+} - \varepsilon_a \partial_z U^{(1)}_{\nu}\bigg|_{0^-} = 0\, ,
\end{eqnarray}
where $\psi$ is the MFT solution for the electrostatic potential obtained in Eq.~(\ref{mf3}),
and the notation $u^{(i)}_{\nu}(\bm{\rho},z)$ and $U^{(i)}_{\nu}({\bf k},z)$, $i=1,2,3$, correspond, respectively,
to the solutions in the three regions: $z<0$ (air, $i=1$), $0 \leq z \leq d$ (proximal layer, $i=2$) and
$z>d$ (distal region, $i=3$). For simplicity, the explicit dependence on ${\bf k}$ is suppressed hereafter, and
a second boundary condition is given in terms of
the macroscopic system size $L$, and is written as $\partial_z u_\nu(\pm L) \to 0$. Note that in the air,
the solution can be obtained by simply taking $\kd=0$ in Eq.~(\ref{ol3}). This leads to an exponentially decaying solution in the air $\sim \exp(\sqrt{k^2+\nu}z)$, $\partial_z U^{(1)}_{\nu}=\sqrt{k^2+\nu}\,U^{(1)}_{\nu}$ at the $z\to 0^{-}$ boundary,
and $k \equiv |{\bf k}|$.

The boundary conditions can then be written in a matrix form
\begin{eqnarray}
\label{ol5}
A{\bf V}\Big|_{z=0} + B{\bf V}\Big|_{z=d} + C{\bf V}\Big|_{z=L} = 0 \, ,
\end{eqnarray}
where the four vector ${\bf V} = \left( U^{(2)}_{\nu}  \, , \,  \partial_z U^{(2)}_{\nu}  \, , \,  U^{(3)}_{\nu} \, , \,  \partial_z U^{(3)}_{\nu} \right)$
and the $4 \times 4$ coefficient matrices $A$, $B$ and $C$ are detailed in Appendix~A.

The boundary conditions are satisfied when the determinant of the coefficient matrix of Eq.~(\ref{ol5}) vanishes. This determinant, $D_\nu({\bf k})$, is called the {\it secular determinant} and can be written as~\cite{fdet}
\begin{eqnarray}
\label{ol9}
D_\nu({\bf k}) = \det\left[A + B \, \Gamma_\nu(a) + C \, \Gamma_\nu(L) \right] \, ,
\end{eqnarray}
where the $4 \times 4$ matrix $\Gamma_\nu(z)$ is also detailed in Appendix~A in terms of
the two independent solutions of the eigenvalue equation (Eq.~(\ref{ol3})).
These two independent solutions are denoted by $h^{(i)}_{\nu}(z)$ and $g^{(i)}_{\nu}(z)$ with $i=2,3$ for regions `2' (proximal) and `3' (distal), respectively.

As was mentioned above, the one-loop fluctuation contribution requires to calculate
${\rm Tr}\ln H_2$. We rely on previous results~\cite{fdet,attard}
where it has been shown that only the $\nu=0$ value
of the secular determinant, $D\equiv D_{\nu=0}$, needs to be evaluated.
This represents an enormous simplification as there is no need to find the entire eigenvalues spectrum of the Hessian.
Let us stress that $\nu=0$  is {\it not} an eigenvalue of the Hessian.

The fluctuation contribution $\Omega_1$ around the MFT can then be written as~\cite{podgornik1989,attard}
\begin{eqnarray}
\label{ol10}
\Omega_1 &=& \half\kbt\, {\rm Tr}\ln\left(H_2({\bf r,r'}) \right) = \nonumber\\
& =& \frac{A\kbt}{8\pi^2} \int\D^2{\bf k}\,\ln\left(\frac{D({\bf k})}{D_{\rm free}({\bf k})}\right) \, ,
\end{eqnarray}
where the integrand depends on the ratio $D({\bf k})/D_{\rm free}({\bf k})$,
and $D_{\rm free}$ is the reference secular determinant for a `free' system without ions.

The next step is to calculate $D({\bf k})$.
We return to Eq.~(\ref{ol3}) to find the solution for $\nu = 0$,
$U_0(z)\equiv U_{\nu = 0}({\bf k},z)$, dropping for convenience the ${\bf k}$-dependence.
Using $\beta e\psi \ll 1$ and keeping only terms of order
$O\left( \e^{-\beta\alpha_{-}} - \e^{-\beta\alpha_{+}}\right)$, Eq.~(\ref{ol3}) yields
\begin{eqnarray}
\label{ol11}
\nonumber& &\Bigg( \frac{\partial}{\partial z}\varepsilon(z)\frac{\partial}{\partial z} -  \varepsilon(z)k^2-  \varepsilon_w\kd^2 \left[ \theta(z-d) - \theta(z-L) \right] \Bigg. \\
& &\Bigg. - \half \varepsilon_w\xi^2 \left[ \theta(z) - \theta(z-d) \right]   \Bigg)U_0(z)=0  \, .
\end{eqnarray}

The four independent solutions of Eq.~(\ref{ol11}), $h_i(z)\equiv h^{(i)}_{\nu=0}(z)$ and
$g_{i}(z)\equiv g^{(i)}_{\nu=0}(z)$, $i=2,3$,
can be written as:
\begin{eqnarray}
\label{ol12}
\nonumber& &h_2 = \cosh qz \quad ; \quad g_2 = \frac{\sinh qz}{q} \\
& &h_3 = \cosh pz \quad ; \quad g_3 = \frac{\sinh pz}{p} \, ,
\end{eqnarray}
where $p^2=k^2+\kd^2$, $q^2=k^2+\xi^2$ and $U_{0}^{(i)}$ is a linear combination of $h_i$ and $g_i$.

By substituting Eq.~(\ref{ol12}) into Eq.~(\ref{ol9}), it is straightforward to compute the secular
determinant which gives,
in the thermodynamical limit, $L \gg d$,
\begin{eqnarray}
\label{ol14}
\nonumber& D(k) \simeq& \Bigg[ \varepsilon_a k\Big( \frac{p}{q}\sinh qd + \cosh qd \Big) \Bigg. \\
& & \Bigg. + \varepsilon_w p\Big( \cosh qd + \frac{q}{p}\sinh qd  \Big) \Bigg]\, \e^{p \,(L-d)}\, .
\end{eqnarray}

The secular determinant can be written in a more familiar way as~\cite{foot1},
%
%%%%%
%
\begin{eqnarray}
\label{ol15}
&D(k) \simeq& \e^{pL} \e^{(q-p)d}\left(\frac{\varepsilon_w q + \varepsilon_a k}{q} \right) \\
\nonumber& &\times\Bigg[ q\Big( 1 - \Delta(q,k) \e^{-2qd} \Big) + p\Big( 1 + \Delta(q,k) \e^{-2qd} \Big) \Bigg] \, ,
\end{eqnarray}
with $\Delta(q,k)$ defined as,

\begin{eqnarray}
\label{ol16}
\Delta(q,k) \equiv \frac{\varepsilon_w q - \varepsilon_a k}{\varepsilon_w q + \varepsilon_a k} \, .
\end{eqnarray}

Keeping linear terms in $d$, the secular determinant from Eq.~(\ref{ol15}) yields,
\begin{eqnarray}
\label{ol18}
D({k}) \simeq \left[\varepsilon_a k + \varepsilon_w p + \varepsilon_w d\left(\xi^2-\kd^2\right)\right]\e^{pL} \, .
\end{eqnarray}
Note that the exact DH result, as obtained in Ref.~\onlinecite{EPL}, is recovered in the above equation for $d=a$.

%%%%%%%%%%%%%%%%%%%%%%%%%%%%%%%%%%%%%%%%%%%%%%%%%%%
\section{Free Energies}
%%%%%%%%%%%%%%%%%%%%%%%%%%%%%%%%%%%%%%%%%%%%%%%%%%%%
The grand potential $\Omega$ can be calculated to the one-loop order by
inserting $D$ obtained in Eq.~(\ref{ol14}) into Eq.~(\ref{ol10}), and expressing $\kd$ of Eq.~(\ref{mf2b})
in terms of the fugacities instead of the bulk densities:
\begin{eqnarray}
\label{f1}
& &\Omega = \Omega_0 + \frac{V\kbt}{12\pi}\left[\left(\Lambda^2+\kd^2\right)^{3/2} - \kd^3 - \Lambda^3 \right] \\
\nonumber& &+ \frac{Ad\kbt}{12\pi}\left[\left(\Lambda^2+\xi^2\right)^{3/2} -
\left(\Lambda^2+\kd^2\right)^{3/2} -\xi^3 + \kd^3 \right] \\
\nonumber& &+~ \frac{A\kbt}{4\pi}\int_0^{\Lambda}\D k\, k \Bigg(\ln \left[\frac{\varepsilon_w q +
 \varepsilon_a k}{2\left(\varepsilon_w + \varepsilon_a \right)k q} \right] \\
\nonumber& &+ \ln \Big[ q\left(1-\Delta(q,k)\e^{-2qd}\right) + \left(1+\Delta(q,k)\e^{-2qd}\right)  \Big] \Bigg)\, ,
\end{eqnarray}
where $\Lambda$ is the ultraviolet (UV) cutoff.
As shown in Appendix~C, for symmetric electrolytes, $\lambda_\pm=n^{(\pm)}_{b}=n_b$.
This simplification is exact for the one-loop order of the  free energy, $F$,
but not for the grand potential, $\Omega$.
In order to find the one-loop grand potential one
has to find consistently the one-loop correction to the fugacities.

Note that the integral in Eq.~(\ref{f1}) has an UV divergency from the upper bound of the  $k$-integral,
as $\Lambda \to \infty$.
Although Coulombic interactions between point-like ions diverge at zero distance,
such a divergence is avoided because of steric repulsion for ions of finite size.
A common way in field theory to avoid this issue without introducing explicitly yet another
steric repulsive interactions, is to employ a short length cutoff.
For isotropic two-dimensional integrals,
as in Eq.~(\ref{f1}) above, the UV cutoff is taken to be $\Lambda = 2\sqrt{\pi}/a$,
where $a$ is the average minimal distance of approach between ions.
This distance can be approximated by $a \simeq 2r$, with $r$ being the ionic radius.
An alternative way of avoiding the divergence is to
use the self-energy regulating technique as in Ref.~\onlinecite{wang2010}.
However, for simplicity  we will employ only to the UV cutoff hereafter.

In order to  calculate the surface tension, we
now calculate the free-energy~\cite{foot2}, which is related to the grand-potential by,
\begin{eqnarray}
\label{f2}
{\cal F} &=& \Omega + \sum_{i\,=\,\pm} \int\D^3r \, \mu_i({\bf r}) n_i({\bf r}) \, .
\end{eqnarray}
It is instructive for the surface tension calculation to separate the volume and surface contributions
of the free energy, ${\cal F}={\cal F}_V+{\cal F}_A$. Taking the $\Lambda\rightarrow\infty$ limit in the
volume term of Eq.~(\ref{f1}), and using Eq.~(\ref{m7}) for $\mu_i$,
yields an expression for ${\cal F}_V$ to the one-loop order:
\begin{eqnarray}
\label{f3}
\nonumber \frac{{\cal F}_V}{V} & &\simeq \frac{\Omega_0}{V} + 2\kbt n_b\ln (n_b a^3) - \frac{\kbt}{12\pi}\kd^3  \\
\nonumber& & -\,\frac{d}{L}\frac{\kbt}{12\pi}\left( \xi^3 - \kd^3\right)
+\,\frac{\kbt}{8\pi}\Lambda\left[ \kd^2 + \frac{d}{L}\left(\xi^2 - \kd^2\right) \right] \\
& &-\, e^2n_b u_{\rm self}({\bf r,r}) \left( 1 + \frac{d}{L}\frac{\xi^2-\kd^2}{\kd^2} \right) \,.
\end{eqnarray}
Note that in the above equation we neglect all terms of order $O(\Lambda^{-1})$.
The first term is the MFT grand potential, $\Omega_0$, the second one is
the usual entropy contribution, and the third one is the well-known volume fluctuation term, as in the DH theory~\cite{Debye1923}.  The fourth and fifth
terms are the bulk self-energies of the ions (diverging with the UV cutoff) and will be shown to  cancel exactly each other.

The surface free-energy ${\cal F}_A$ to one-loop order is:
\begin{eqnarray}
\label{f4}
\frac{{\cal F}_A}{A}& &=  \frac{\kbt}{4\pi}\int_0^{\Lambda}\D k\, k \Bigg(\ln \left[\frac{\varepsilon_w q +
\varepsilon_a k}{2\left(\varepsilon_w + \varepsilon_a \right)k q} \right] \\
\nonumber& &+ \ln \Big[ q\left(1-\Delta(q,k)\e^{-2qd}\right) + p\left(1+\Delta(q,k)\e^{-2qd}\right)  \Big] \Bigg) \, .
\end{eqnarray}

We now treat the spurious divergencies that originate from the bulk self-energies of the ions.
The Coulomb potential obeys:
\begin{eqnarray}
\label{u1}
&\nabla^2u({\bf r},{\bf r}')=-\frac{4\pi}{\varepsilon_w}\delta({\bf r}-{\bf r}')  \qquad &{\rm for} \quad z > 0 \, ,\nonumber\\
&\nabla^2u({\bf r},{\bf r}')=0 \qquad &{\rm for} \quad z < 0 \, .
\end{eqnarray}
As our system exhibits translational invariance in the transverse $(x,y)$ directions,
we can simplify the transverse Fourier transform of Eq.~(\ref{ol2}) into the Fourier-Bessel representation
\begin{eqnarray}
\label{u2}
u({\bf r},{\bf r}') &=& \frac{1}{4\pi^2}  \int d^2{\bf k}  ~U_0(k; z,z') \e^{i{\bf k} \cdot \bm{\rho}} = \nonumber\\
& & = \int dk\, k\,  U_0(k; z,z') J_0(k \vert \bm{\rho} - \bm{\rho}'\vert) \, ,
\end{eqnarray}
where $J_0(z)$ is the zeroth-order Bessel function of the 1st kind,
and $\bm\rho = (x,y)$ is the in-plane radial vector.
The solution of Eq.~(\ref{u1})  is
\begin{eqnarray}
\label{u3}
U_0(k; z,z') = \frac{2\pi}{\varepsilon_w k}\Big( \e^{-k\vert z - z'\vert} + \frac{\varepsilon_w-
\varepsilon_a}{\varepsilon_w+\varepsilon_a}\e^{-k (z+z')}\Big)\, , \quad
\end{eqnarray}
for $z>0$, and
\begin{eqnarray}
\label{u3a}
U_0(k; z,z') = \frac{2\pi}{\varepsilon_w k}\Big( 1 + \frac{\varepsilon_w-\varepsilon_a}{\varepsilon_w+\varepsilon_a}\Big)\e^{k (z-z')}\, ,
\end{eqnarray}
for $z<0$.
The self-energy of an ion in the bulk is obtained by setting ${\bf r} = {\bf r}'$ and $z\to\infty$
\begin{eqnarray}
\label{u4}
\nonumber u_{\rm self}({\bf r},{\bf r}) = \frac{1}{\varepsilon_w } \int_0^{\Lambda}\!\! \D k
\left(1+\frac{\varepsilon_w-\varepsilon_a}{\varepsilon_w+\varepsilon_a} \e^{-2 k z} \right) \simeq \Lambda/\varepsilon_w\, . \\
\end{eqnarray}
The integral over $k$ for the self-energy, Eq.~(\ref{u4}),
has a UV cutoff that was replaced by its most divergent term,
which is linear in $\Lambda$. Finally, by substituting Eq.~(\ref{u4}) into Eq.~(\ref{f3}),
the two diverging terms in  $F_V$ cancels each other.

To summarize, we have shown that within our approach the
self-energy diverging terms cancel out and, as anticipated, they do not affect the free energy.

%%%%%%%%%%%%%%%% fig 2%%%%%%%%%%%%%%%%%%%%
%Fig 2
\begin{figure*}[th]
\center
\includegraphics[scale=0.55,draft=false]{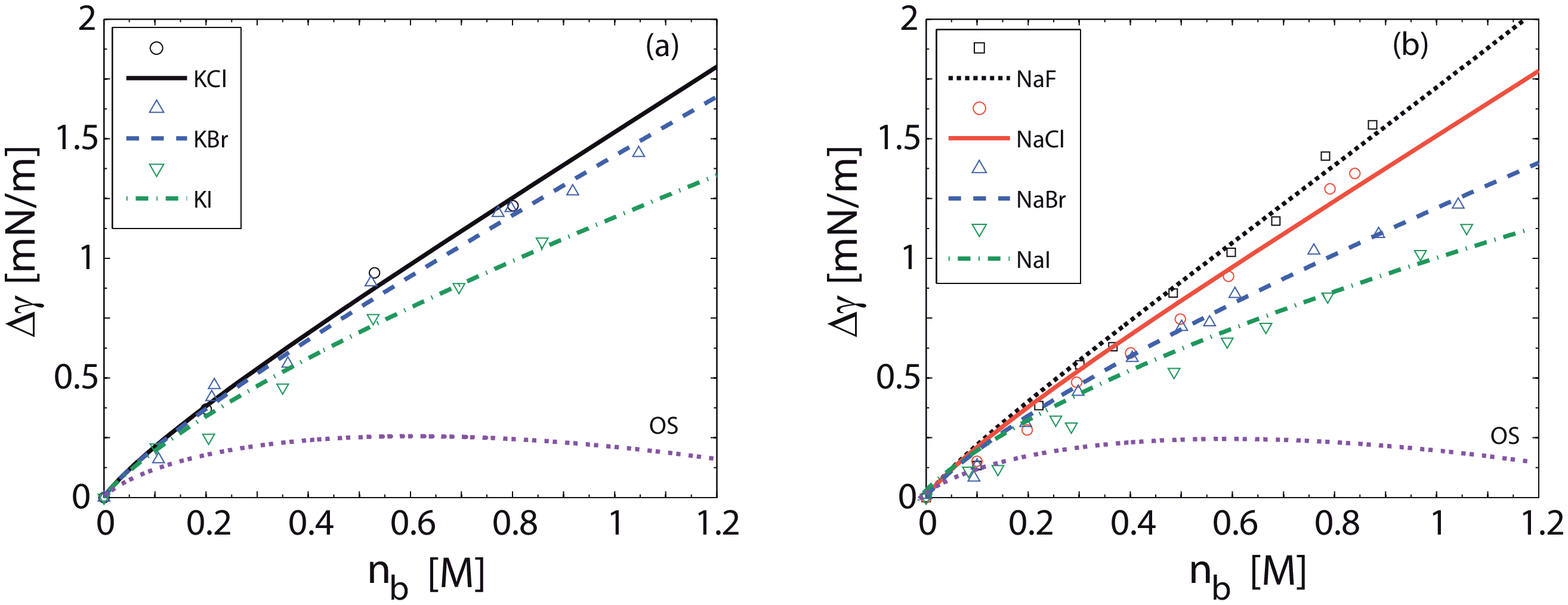}
\caption{\textsf{(color online).
Comparison of the fitted excess surface tension, $\Delta\gamma$, at the air/water interface
with experimental data from Refs.~\onlinecite{KExp} (For the K$^+$
series) and~\onlinecite{NaExp} (for the Na$^+$ series), as function of ionic concentration, $n_b$. For (a) ${\rm KX}$ and for (b) ${\rm NaX}$,
where ${\rm X}$ stands for one of the Halogen anions ${\rm F^-}$, ${\rm Cl^-}$, ${\rm Br^-}$, and ${\rm I^-}$.
The bottom dashed line represents the OS surface tension~\cite{onsager_samaras}.
The fitted adhesivity values, $\alpha^{*}$ in units of $k_B T$, are shown in Tables~I and II.
Other parameters are $T=300$\,K, $\varepsilon_w=80$ (water) and $\varepsilon_a=1$ (air).
}}\label{fig2}
\end{figure*}
%%%%%%%%%%%%%%%%%%%%%%%%%%%%%%%%%%%%%%%%%%%

%%%%%%%%%%%%%%%%%%%%%%%%%%%%%%%%%%%%%%%%%%%%%%%%%%%
\section{Surface Tension}
%%%%%%%%%%%%%%%%%%%%%%%%%%%%%%%%%%%%%%%%%%%%%%%%%%%%

We are interested in calculating the surface tension of an aqueous ionic solution. The excess surface
tension, $\Delta\gamma$, can be obtained from the Gibbs absorbtion isotherm or, equivalently,
by taking the difference between the
free-energy of the two-phase system and the sum of the free energies
of the two semi-infinite systems of the air and the bulk ionic solution:
\begin{eqnarray}
\label{st1}
\Delta\gamma&= &\left[{\cal F}(2L) - {\cal F}^{\rm (air)}(L)- {\cal F}^{\rm (B)}(L)\right]/A \, .
\end{eqnarray}
Here, $\Delta\gamma=\gamma-\gamma_{\rm A/W}$ is the excess ionic contribution to the
surface tension with respect to the surface tension between pure water and air, $\gamma_{\rm A/W}$.

The bulk electrolyte free energy, $F^{\rm (B)}(2L)$,
is obtained from Eqs.~(\ref{f3}) and (\ref{f4})
by considering the entire $[-L,L]$ interval as a uniform dielectric medium with
$\varepsilon_w$, $\alpha_{\pm}=0$ and mobile ions. This implies $\varepsilon_a\to \varepsilon_w$ and $q \to p$.
Then, $D({\bf k})/D_{\rm free}({\bf k})=p/k$,
while $\Omega_0=-2n_b V$ is obtained for the MFT solution of $\psi=0$ (bulk electrolyte phase).
Applying these limiting values gives the  free-energy of a system of width $2L$. This system includes two boundaries
at the two extremities, $\pm L$. Thus, one needs to divide the free energy by two in order to get the bulk free-energy of a slab of width $L$:
\begin{eqnarray}
\label{st2}
\nonumber&{\cal F}^{\rm (B)}(L) &= \kbt V\left[2n_b\ln \left( n_b a^3\right) - 2n_b - \frac{\kd^3}{12\pi}\right] \\
& &+ \frac{A\kbt}{8\pi }\int_0^{\Lambda}\D k \,k \ln\left(\frac{p}{k}\right) \, .
\end{eqnarray}
The free-energy of the air phase is equal to zero, ${\cal F}^{\rm (air)}(L) = 0$,
as there are no ions in the air phase.
For calculating the interfacial tension between water and oil,
for typical oil with dielectric constant $\varepsilon_a \simeq 2$,
the assumption that no ions penetrate into the oil phase will be maintained,
and the free energy of the oil phase will be taken as zero.

We calculate in the following subsections the excess surface tension to one-loop order, $\Delta\gamma =
\Delta\gamma_0 + \Delta\gamma_1$, where $\Delta\gamma_0$ is the MFT contribution to the surface tension,
while $\Delta\gamma_1$ is its one-loop correction.

%%%%%%%%%%%%%%%%%%%%%%%%%%%%%%%%%%%%%%%%%%%%%%%%%%%%
\subsection{Mean-Field Surface Tension}
%%%%%%%%%%%%%%%%%%%%%%%%%%%%%%%%%%%%%%%%%%%%%%%%%

Using the mean-field $\psi(z)$ from Eq.~(\ref{mf2}), the MFT surface tension is
\begin{eqnarray}
\label{sm1}
\nonumber\Delta\gamma_0 &=& -\int_{-\infty}^{\infty}\!\!\!\!\!\D z \frac{\varepsilon(z)}{8\pi}\left(\frac{d\psi}{dz}\right)^2 \\
\nonumber& & +\, \kbt n_b \int_0^d \D z \,  \left(2 - \e^{-\beta e\psi -\beta\alpha_+} - \e^{\beta e\psi -\beta\alpha_-} \right)  \\
& &+\, 2\kbt n_b \int_d^L \D z \, \left(1 - \cosh\beta e\psi\right)\, .
\end{eqnarray}
As the electrostatic potential and its derivative are of the order $O\left(\e^{-\beta\alpha_-} -
\e^{-\beta\alpha_+} \right)$, the MFT surface tension is simply,
\begin{eqnarray}
\label{sm2}
\Delta\gamma_0 \simeq -\kbt n_b d \left[ \left(\e^{-\beta\alpha_-} -1 \right) + \left( \e^{-\beta\alpha_+} -1 \right) \right] \, .
\end{eqnarray}
On the MFT level, the equation above is exact for $\alpha_{-} = \alpha_{+}$.
To show this, we first substitute $\alpha_{-} = \alpha_{+}$
in Eq.~(\ref{mf3}), leading to vanishing electric field and potential.
Further substitution of $\psi=0$ and $\partial_z\psi = 0$ in Eq.~(\ref{sm1}) leads to Eq.~(\ref{sm2})
without any further approximations.

In order to make contact with the PB surface tension
of a charged surface with adhesivity, we write Eq.~(\ref{sm1}) to linear order in $d$.
Substituting the first integration of Eq.~(\ref{mf1}) into Eq.~(\ref{sm1}), one gets
\begin{eqnarray}
\label{sm3}
\nonumber\beta \Delta\gamma_0 &=& - n_bd \Big[ \left(\e^{-\beta\alpha_-} -1 \right)\e^{\beta e\psi} + \left( \e^{-\beta\alpha_+} -1 \right)\e^{-\beta e\psi} \Big]   \\
&-& 8  n_b\kd^{-1}\Big(\cosh(\beta e\psi_s/2) - 1\Big)\, .
\end{eqnarray}
The second term in the above equation is the PB surface tension. For our purposes, the electrostatic potential is small
and the same second term is negligible. In this case, the above equation exactly coincides with Eq.~(\ref{sm2}).

%%%%%%%%%%%%%%%%%%%%%%%%%%%%%%%%%%%%%%%%%%%%%%%%%%%%
\subsection{One-loop Correction to the Surface Tension}
%%%%%%%%%%%%%%%%%%%%%%%%%%%%%%%%%%%%%%%%%%%%%%%%%%%%

The one-loop correction to the surface tension, $\Delta\gamma_1$, is obtained from Eq.~(\ref{f4}):
\begin{eqnarray}
\label{so1}
\nonumber& &\frac{8\pi}{\kbt} \Delta\gamma_1 = \int_0^{\Lambda}\D k\, k \Bigg(\ln \left[\frac{k}{p}
\left(\frac{\varepsilon_w q + \varepsilon_a k}{2\left(\varepsilon_w + \varepsilon_a \right)k q} \right)^2\right] \\
\nonumber& &+ \ln \Big[ q\left(1-\Delta(q,k)\e^{-2qd}\right) + p\left(1+\Delta(q,k)\e^{-2qd}\right)  \Big]^2 \Bigg) \\
& &- \,\frac{2d}{3}\left( \xi^3 - \kd^3\right) \, .
\end{eqnarray}
Taking the limit of $d\to0$ or $\alpha_{\pm} \to 0$ (which implies $q\to p$), gives the OS result
with a correction due to the finite ion size~\cite{podgornik1988}. By keeping terms linear in $d$,
the linearized fluctuation contribution yields~\cite{foot3}:
\begin{eqnarray}
\label{so2a}
& &\frac{8\pi}{\kbt}\Delta\gamma_1 \simeq \int_0^{\Lambda}\D k \,k  \ln\left[\frac{k}{p}
\left(\frac{ \varepsilon_a k +\varepsilon_w p }{k\left(\varepsilon_w + \varepsilon_a \right)}\right)^2 \right] \\
\nonumber& &\,\,+\, 2d\left[ \int_0^{\Lambda}\D k \,k\left(q-p\right)\frac{\varepsilon_w q - \varepsilon_a k}{ \varepsilon_w p + \varepsilon_a k }
- \frac{\xi^3 - \kd^3}{3} \right] \, .
\end{eqnarray}
The above result contains the OS result~\cite{onsager_samaras,podgornik1988,dean2004} and
an ionic-specific correction, as will be discussed later.
It is clear from the above results that as long as the adhesivity parameters $\alpha_{\pm}$ are
small, the MFT contribution to the surface tension, $\Delta\gamma_{0}$, is small and the dominant
contribution comes from the fluctuation term, $\Delta\gamma_{1}$. This observation goes hand in
hand with the fact that the OS result by itself originates from fluctuations beyond MFT.

The leading asymptotic behavior of the integral of Eq.~(\ref{so1}) reveals the OS result and its correction terms.
Writing down only the remaining $\Lambda$-dependent terms, we obtain
\begin{eqnarray}
\label{so3}
\nonumber \frac{8\pi}{\kbt}\Delta\gamma_{1} &\simeq& -
\left(\frac{\varepsilon_{w}-\varepsilon_a}
{\varepsilon_w+\varepsilon_a}\right)\frac{\kappa_D^2}{2} \left[ \ln\left(\frac{1}{2}\kd\lb\right)
- \ln\left(\frac{1}{2}\lb\Lambda \right)\right. \nonumber\\
&-&  \left. \frac{2\left[ \varepsilon_w d\left(\xi^2-\kd^2\right) \right]^2}{\kappa_D^2(\varepsilon_w^2-
\varepsilon_a^2)} \ln\left(\kappa_D\Lambda^{-1}\right) \right] \, ,
\end{eqnarray}
with the Bjerrum length defined as $\lb = \beta e^2/\varepsilon_w$ and  $\xi^2$ defined in Eq.~(\ref{xidef}).
with $\xi^2$ defined in Eq.~(\ref{xidef}).
The first term in $\Delta\gamma_{1}$  is the well-known OS result~\cite{onsager_samaras,podgornik1988,dean2004}
and it varies as $\sim\kappa_D^2\ln(\kappa_D\lb)$, the second term is a correction due to the ion minimal distance
of approach with $\Lambda=2\sqrt{\pi}/a$, while the third term is a correction related to the adhesivity parameters,
$\alpha_{\pm}$, through $\xi^2 \equiv \kd^2\left( \e^{-\beta\alpha_{-}} + \e^{-\beta\alpha_{+}}\right)/2 $.
For $\beta\alpha_{\pm} \ll 1$, the latter term is negligible and the derived surface tension agrees well with
the OS result, as expected.

%%%%%%%%%%%%%%%% fig 3 %%%%%%%%%%%%%%%%%%%%
%Fig 3
\begin{figure}[bh]
\center
\includegraphics[scale=0.6,draft=false]{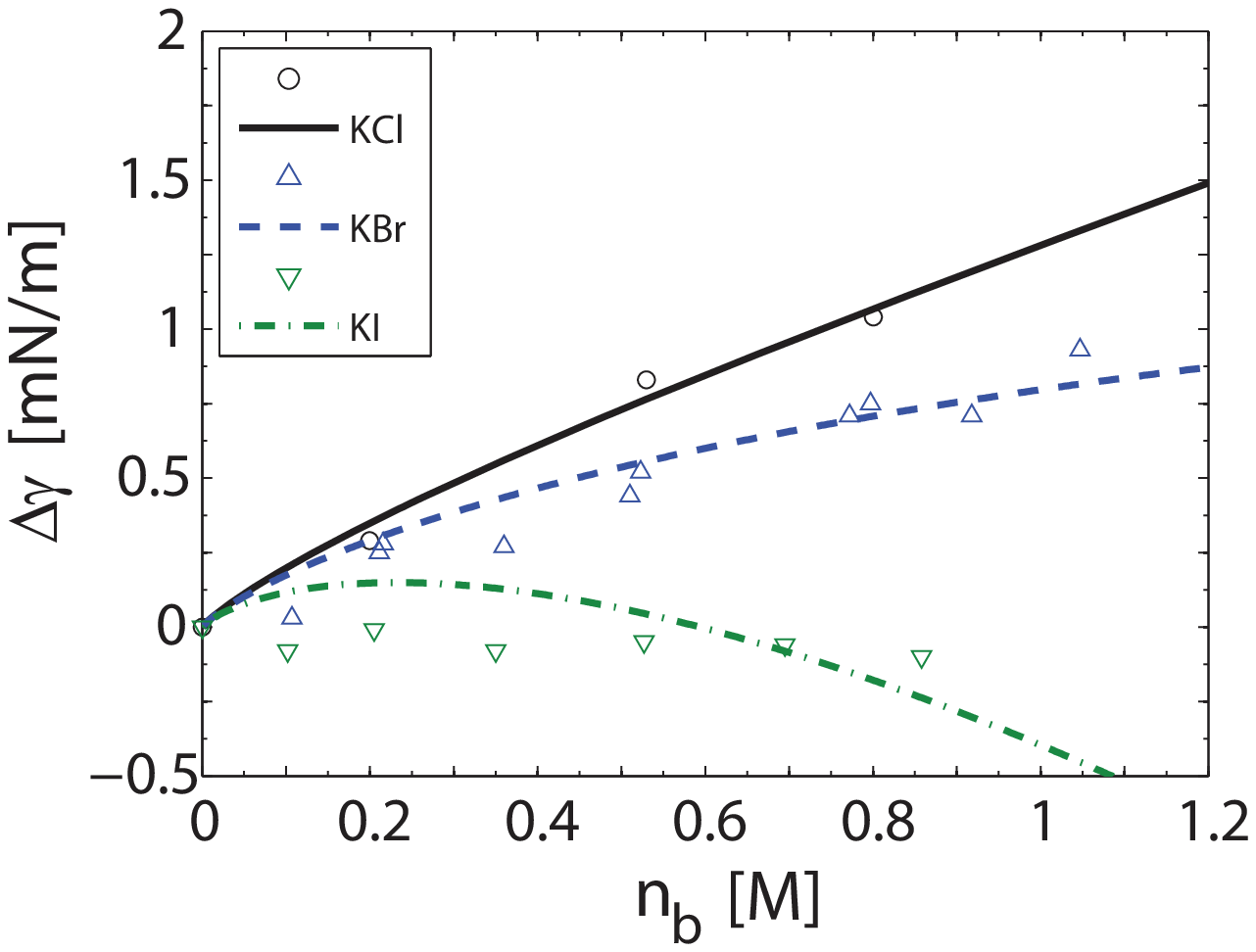}
\caption{\textsf{(color online).
Comparison of the fitted excess surface tension for ${\rm KX}$ electrolytes, where
${\rm X}$ stands for one of the Halogen anions ${\rm Cl^-}$, ${\rm Br^-}$ and ${\rm I^-}$,
with experimental data from Ref.~\onlinecite{KExp},
as function of ionic concentration, $n_b$, for dodecane/water.
The fitted adhesivity values, $\alpha^{*}$, are shown in Table~I.
All other parameters are as in Fig.~\ref{fig2}, beside the dielectric constant of dodecane, $\varepsilon_d=2$.}}
\label{fig3}
\end{figure}
%%%%%%%%%%%%%%%%%%%%%%%%%%%%%%%%%%%%%%%%%%%%%

For salt concentration larger than $\sim 0.3\kbt$,
the analytical approximation shown in Eq.~(\ref{so3})
deviates from the full numerical solution of Eq.~(\ref{so2a}) .
It is nevertheless possible to solve Eq.~(\ref{so2a}) analytically.
Along these lines, we present in Appendix~B an analytical solution to order $O(1/\Lambda)$, Eq.~(\ref{b1}),
where we do not neglect constants (with respect to $\Lambda$).
This solution is almost equivalent to the full numerical solution of Eq.~(\ref{so2a}).

\begin{table*} [th]
\begin{tabular}{|             c                                |}
\hline
\qquad\qquad\qquad\qquad\qquad Air  \qquad\qquad\qquad\qquad\,\,\,\,\,\,\,\,\,\,\,\,\,\,\,  \vline\,\,\vline
\,\qquad\qquad\qquad\qquad  Oil \qquad\qquad\qquad\qquad\qquad\,\,\, \\
\end{tabular} \\
\begin{tabular}{| c | c | c | c | c | c | c | c | c | c | c | c |}
  \hline
  & \,\, $d \, [{\rm \AA}]$  \,\,  &  \,\,  $\alpha^{*} \, [\kbt]$  \,\, & \,\, $\alpha_{-}\, [\kbt]$ \,\, & \,\,
  $\alpha_{+}\, [\kbt]$ \,\, &\quad& \,\, $d \, [{\rm \AA}]$  \,\,  &  \,\,  $\alpha^{*} \, [\kbt]$  \,\, & \,\,
  $\alpha_{-}\, [\kbt]$ \,\, & \,\, $\alpha_{+}\, [\kbt]$ \,\, \\ \hline

  %NaF  &  7.1  & 0.2813 & 0.1515  & 0.1111 &\quad&-&-&-&-\\
  %NaCl &  6.9  & 0.2052 & 0.0851  & 0.1111 &\quad&-&-&-&-\\
  %NaBr &  6.88 & 0.0991 & -0.0101 & 0.1111 &\quad&-&-&-&-\\
  %NaI  &  6.89 & 0.0330 & -0.0707 & 0.1111 &\quad&-&-&-&-\\ \hline

  KCl &  6.63 & 0.2112 & 0.0851  & 0.1652 &\quad& 6.63 & 0.1171  & -0.0050 & 0.1313\\
  KBr &  6.61 & 0.1732 & -0.0101 & 0.1652 &\quad& 6.61 & -0.0390 & -0.1451 & 0.1313\\
  KI  &  6.62 & 0.0811 & -0.0707 & 0.1652 &\quad& 6.62 & -0.3333 & -0.4141 & 0.1313\\ \hline
\end{tabular}
\caption{\textsf{Fitted values for $\alpha^{*}$ and $\alpha_{\pm}$ for the air/water and oil/water interfaces.
The $\alpha_{\pm}$ are obtained by the procedure elaborated in the text and include also predictions for ${\rm KCl}$ and ${\rm KBr}$ at the air/water interface.
The ion radii are obtained from Ref.~\onlinecite{IonRadii}.    }}\label{table1}
\end{table*}

%%%%%%%%%%%%%%%%%%%%%%%%%%%%%%%%%%%%%%%%
\section{Comparison to Experiments}
%%%%%%%%%%%%%%%%%%%%%%%%%%%%%%%%%%%%%%%%

We now compare our results for the surface tension,
$\Delta\gamma = \Delta\gamma_0 + \Delta\gamma_1$, with experimental data.
The surface tension prediction is obtained from $\Delta\gamma_0$ in Eq.~(\ref{sm2}) and the numerical solution,
$\Delta\gamma_1$  of Eq.~(\ref{so1}).

For simplicity, we take $d$, the thickness of the proximal layer,
to be equal to $a$, the average minimal distance between cations and anions,
$a = r_{+}^{\rm hyd} + r_{-}^{\rm hyd}$.
Furthermore, we treat $\alpha_{\pm}$ as fit parameters.
We note that the obtained results for the surface tension, $\Delta\gamma =  \Delta\gamma_0 +
\Delta\gamma_1$, are symmetric with respect to $anion \longleftrightarrow cation$ exchange.
This is important because it means that two parameter fit with
$\alpha_{\pm}$ will always give two equivalent results, $\alpha_{+}\leftrightarrow\alpha_{-}$.
Furthermore, for $\beta\alpha_{\pm} \ll 1$ one can define
$\alpha^{*} = \alpha_- + \alpha_+$ as a single fit parameter, and the fit with $\alpha^{*}$
produces almost equivalent results to the fit with the two independent parameters.

We start by presenting the fits of the experimental data with $\alpha^{*}$.
In Fig.~\ref{fig2} we compare our theory at the air/water interface for
(a) three different ionic solutions with K$^+$ as their cation,
and (b) four different ionic solutions with Na$^+$ as their cation.
The fits in (a) are in very good agreement with experiments.
In (b), for the larger Br$^{-}$ and I$^{-}$ anions (with respect to their crystallographic size),
the fit agrees well for the entire concentration range up to $\sim$1\,M,
while for the smaller F$^-$ and Cl$^-$ anions,
some deviations at concentrations larger than $\sim0.8$\,M can be seen.

Our model can also be applied successfully to other types of liquid interfaces, such as oil/water, and to
 more complex anions such as {\it oxy anions}, which are defined by the generic formula, ${\rm A_xO_y^{-}}$,
 or acids such as ${\rm HCl}$ and ${\rm HClO_4}$.
As hydrogen can form complexes with water molecules,
the ${\rm H^+}$
represents all of these complexes.
We do not need to know the specific complexation of the hydrogen in water,
but only its effective radius~\cite{Marcus}.

\begin{figure}[!h]
\center
\includegraphics[scale=0.60,draft=false]{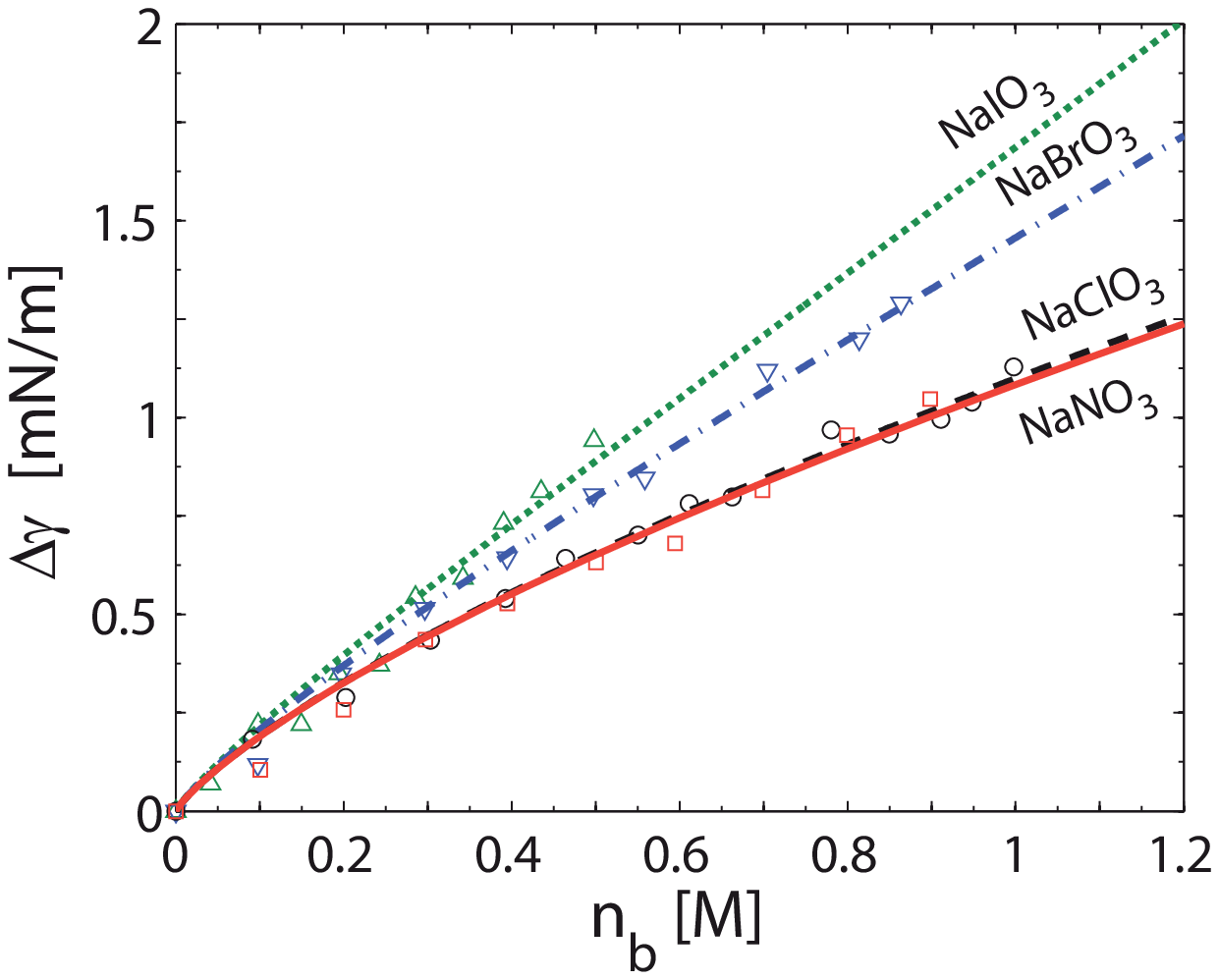}
\caption{\textsf{(color online).
Comparison of the fitted excess surface tension at the air/water interface for ${\rm NaX}$ electrolytes, where
${\rm X = IO_3^- , BrO_3^-, ClO_3^-, NO_3^-}$ stands for one of the oxy anions,
with experimental data from Refs.~\onlinecite{OxyExp,levin2010}, as function of ionic concentration, $n_b$.
The fitted adhesivity values, $\alpha^{*}$ are shown in Table II.
All other parameters are as in Fig.~\ref{fig2}.
}}\label{fig4}
\end{figure}

In Fig.~\ref{fig3}, we compare the fits for oil/water interface,
where in the experiments dodecane is used as the oil.
The fits are done for three different salts having in common the K$^+$ cation.
The fits for ${\rm KCl}$ and ${\rm KBr}$ are in very good agreement with experiments,
while the fit for KI is not as good.
The surface tension of KI shows a very small $\Delta\gamma$,
which is almost independent of the salt concentration and, hence, harder to fit. In Fig.~\ref{fig4} we fitted yet another series of
four different oxy anions (with Na$^+$ as their cation), and the fits in the figure are in excellent agreement with experiments.

%%%%%%%%%%%%%%%% fig 4%%%%%%%%%%%%%%%%%%%%
%%%%%%%%%%%%%%%%%%%%%%%%%%%%%%%%%%%%%%%%%%%

\begin{table} [!h]
\begin{tabular}{| c | c | c | c | c |}
  \hline
  & \,\, $d \, [{\rm \AA}]$  \,\,  &  \,\,  $\alpha^{*} \, [\kbt]$  \,\, & \,\, $\alpha_{-}\, [\kbt]$ \,\, & \,\, $\alpha_{+}\, [\kbt]$ \,\, \\ \hline

  NaF     &  7.1  & 0.2813 & 0.1515  & 0.1111 \\
  NaCl    &  6.9  & 0.2052 & 0.0851  & 0.1111 \\
  NaBr    &  6.88 & 0.0991 & -0.0101 & 0.1111 \\
  NaI     &  6.89 & 0.0330 & -0.0707 & 0.1111 \\ \hline

  ${\rm NaIO_3}$   &  7.32 & 0.2673 & 0.1391   & 0.1111 \\
  ${\rm NaBrO_3}$  &  7.09 & 0.1852 & 0.0671   & 0.1111 \\
  ${\rm NaClO_3}$  &  6.99 & 0.0651 & -0.0410  & 0.1111 \\
  ${\rm NaNO_3}$   &  6.93 & 0.0591 & -0.0470  & 0.1111 \\
  %${\rm NaClO_4}$  &  6.96 & -0.2272& -0.3093  & 0.1111 \\
  \hline

  HCl   &  4.32 & -0.6116 & 0.0851   & -0.6537\\
  ${\rm HClO_4}$ &  4.38 & -0.9898 & -0.5696  & -0.6537\\
  \hline
\end{tabular}
\caption{\textsf{Fitted values for $\alpha^{*}$ and $\alpha_{\pm}$ at the air/water interface.
The $\alpha_{\pm}$ are obtained by the procedure elaborated in the text and include a prediction for ${\rm HClO_4}$ at the air/water interface.
The radii for all ions except ${\rm H^+}$ are obtained from Ref.~\onlinecite{IonRadii}.
The ${\rm H^+}$ effective radius is obtained from Ref.~\onlinecite{Marcus}.    }}
\label{table2}
\end{table}

It is reasonable to assume that for the same interface the adhesivities of the anions/cations will
not differ significantly for different combinations of anion-cation pairs.
For example, we can calculate $\alpha_{\rm KI}^{*}$
even when we use $\alpha^*$ as the only fit parameter.
This can be done with a simple substitution,
$\alpha_{\rm KI}^{*} \simeq \alpha^{*}_{\rm NaI} - \alpha^{*}_{\rm NaBr} + \alpha^{*}_{\rm KBr} \simeq 0.1071$
Using this procedure it is possible to obtain reasonable predictions for $\alpha^{*}$ for additional salts.

We also tried to fit the experimental data with two independent parameters for anions ($\alpha_-$) and cations ($\alpha_+$).
Just taking the ``best fit" is not sufficient for these fits.
The first problem is that our results do not distinguish between anions and cations, as explained earlier.
This implies that the cation and anion adhesivities have to be attributed from other external considerations.

The other, and more significant issue, is the fact that many pairs of $\alpha_{\pm}$ gives similar excellent fits.
We first fit all the electrolytes at the air/water interface in the following way.
(i) We fit ${\rm NaF}$ and choose from the `best fit' the larger adhesivity to be $\alpha_{+}$ as it is
known~\cite{levin200y,collins1985} that ${\rm F^-}$ is more hydrated than ${\rm Na^+}$.
(ii) We then use the ${\rm Na^+}$ adhesivity and fit all other ions in the ${\rm NaX}$ series.
(iii) We continue by using the fitted $\alpha_{-}$ for ${\rm I^-}$ adhesivity and fit the ${\rm KI}$ electrolyte.
(iv) With the ${\rm K^+}$ adhesivity and the adhesivities obtained from the ${\rm NaX}$ fits, we can
make predictions for the air/water surface tension of ${\rm KBr}$ and ${\rm KCl}$.

The same procedure was also applied to two acids, ${\rm HCl}$ and ${\rm HClO_4^-}$ (see Table~II). First, we
used the ${\rm Cl^-}$  adhesivity to fit ${\rm HCl}$,
and then the ${\rm H^+}$ adhesivity to fit ${\rm HClO_4^-}$.
Finally, for the ${\rm K^+}$ fits at the oil/water interface, we first fitted ${\rm KI}$ and then used the
adhesivity of ${\rm K^+}$ to fit all other ${\rm KX}$ salts at the oil/water interface.
The results of these fits are all presented in Tables~I and II.
We do not show these fits in the figures as they are almost identical to the fits in Figs.~2-5,
where a single $\alpha^{*}$ parameter is utilized.

\begin{figure}[h!]
\center
\includegraphics[scale=0.55,draft=false]{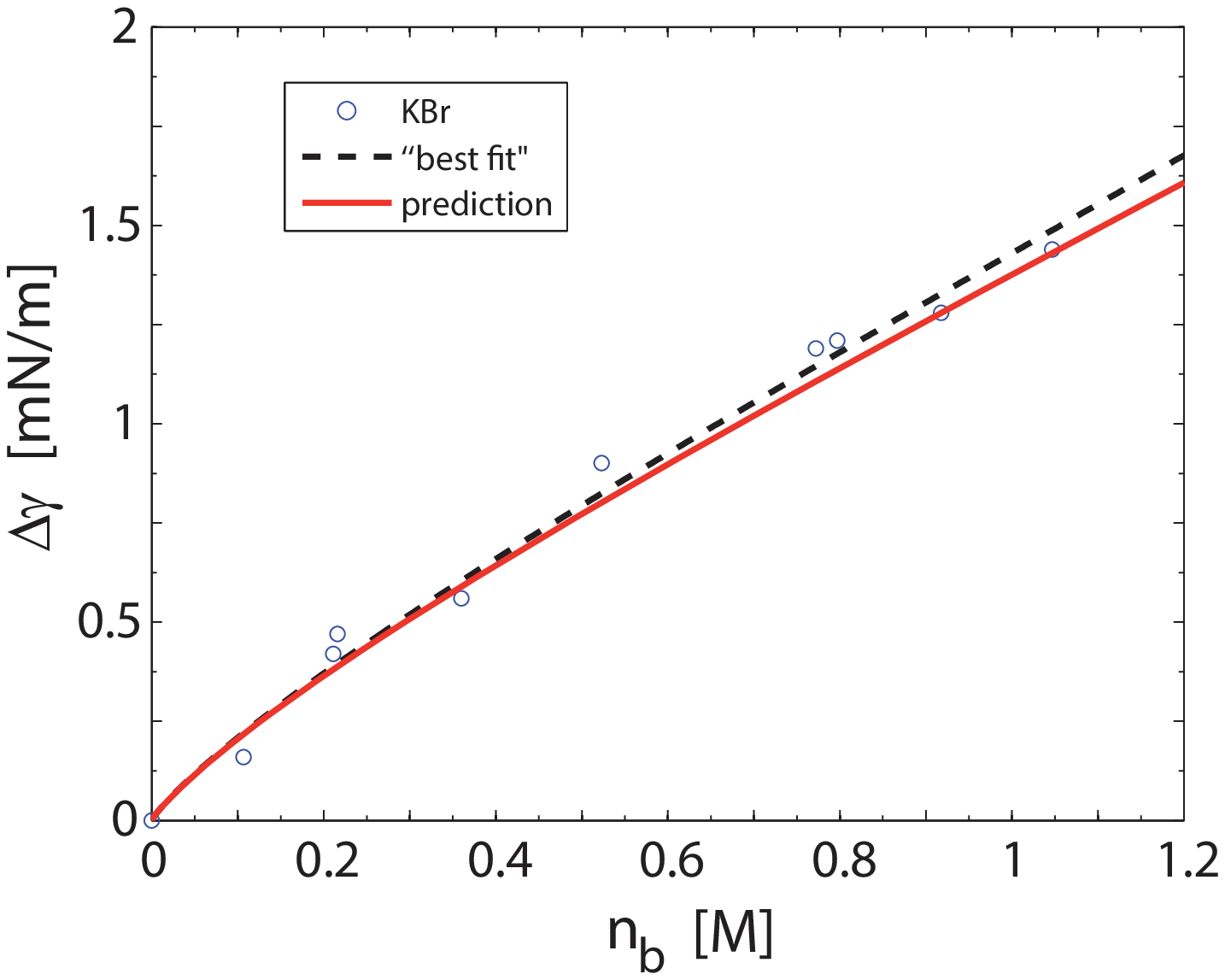}
\caption{\textsf{(color online).
Comparison of the ``best fit'' to the data (dashed line) and the predicted (red solid line)
excess surface tension for ${\rm KBr}$ at the air/water interface as function of ionic concentration, $n_b$.
The fit parameter values, $\alpha_{\pm}$, are shown in Table~I.
}}\label{fig6}
\end{figure}

In Fig.~\ref{fig6} we show the difference of our prediction to ${\rm KBr}$ from the `best fit' of the data.
The predicted surface tension is excellent. The difference from the ``best fit'' is very small, and up to $0.8$\,M it is almost unnoticeable.
We do not show the prediction of ${\rm KCl}$, but it is just as good.
It is important to note that other fitting strategies
would have given different results for $\alpha_{\pm}$, and probably would result in good fits.
Our model gives good estimates of the adhesivity parameters for physical systems,
and we can consistently fit a large number of experiments and even make reasonable predictions.

\begin{figure}[h!]
\center
\includegraphics[scale=0.55,draft=false]{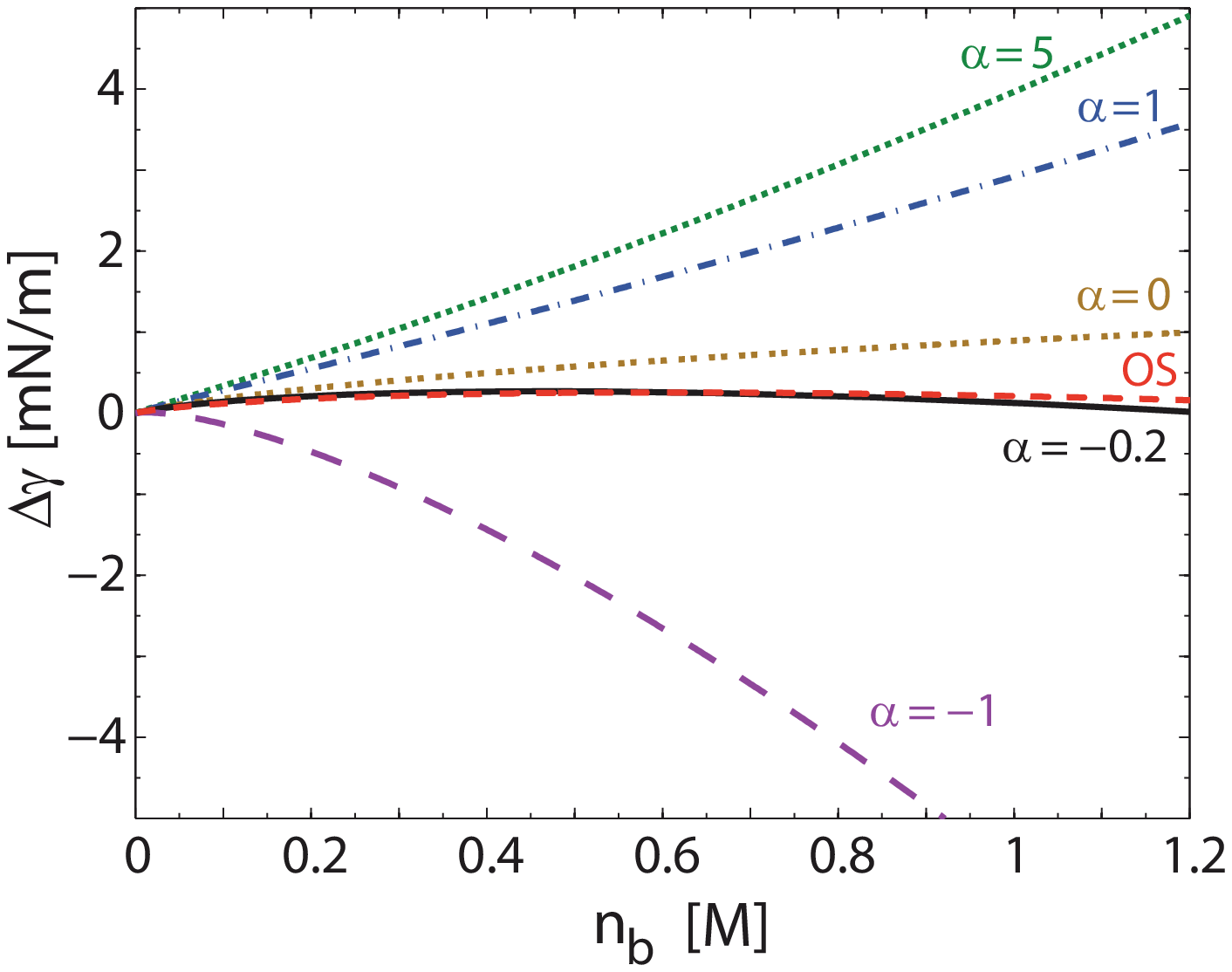}
\caption{\textsf{(color online).
Excess surface tension at the air/water interface as function of ionic concentration,
$n_b$ for different values of $\alpha$ (in units of $k_B T$).
The OS result is obtained for $\alpha \simeq -0.2k_B T$.
All other parameters are as in Fig.~\ref{fig2}.
}}\label{fig8}
\end{figure}

In Fig.~\ref{fig8} we plot the excess surface tension at the air/water interface, $\Delta\gamma$,
the sum of Eqs.~(\ref{sm2}) and (\ref{so1}),
as function of ionic concentration, $n_b$, for different values of $\alpha$.
The OS result is recovered almost exactly for $\alpha \simeq -0.2k_B T$.
Notice that the surface tension has an upper bound for $\alpha \to \infty$,
but this bound is practically reached for $\alpha \gtrsim 5k_B T$.
This happens for infinite repulsion when all the ions are expelled from the proximal layer.
In addition, in order to obtain this bound,
we require that the ionic concentration at the surface cannot be negative.
This is not the case for $\alpha \to -\infty$,
which leads to an infinite amount of anions to be adsorbed on the surface.
This is not physical because the actual upper bound is the
close packing concentration, even for an infinite attraction.
As we did not take into account the finite size of the ions and,
we get this non-physical situation.
In this paper we expect, and obtain, only small values of adhesivities.
Hence, this non-physical situation is not relevant.

%%%%%%%%%%%%%%%% fig 5%%%%%%%%%%%%%%%%%%%%
%%%%%%%%%%%%%%%%%%%%%%%%%%%%%%%%%%%%%%%%%%%

We can now arrange the various ions in an extended reverse Hofmeister series with decreasing adhesivity strength at the air/water interface.
The anions series is:
${\rm F}^{-} > {\rm IO_3}^{-}  > {\rm Cl}^{-} > {\rm BrO_3}^{-} > {\rm Br}^{-} > {\rm ClO_3}^{-} > {\rm NO_3}^{-} > {\rm I}^{-}$,
while for cations it is:
${\rm K}^{+} > {\rm Na}^{+}$.

At the oil/water interface, Fig.~\ref{fig3}, the same reversed Hofmeister series emerges and
the interaction becomes more attractive. This effect is substantially stronger for the anions,
and might be connected with the stronger dispersion forces at the oil/water interface,
related directly to the large anion polarizabilities~\cite{ninham2001}.
We denote the difference in adhesivity between the air/water and oil/water interfaces by $\Delta \alpha=\alpha({\rm a/w})-\alpha({\rm o/w)}$.
From the fitted values of $\alpha^*$, $\Delta\alpha$ is different for each anion, and $\Delta\alpha_{\rm I} > \Delta\alpha_{\rm Br} > \Delta\alpha_{\rm Cl}$.
This can be explained by a change in the water-surface interaction.
If the water-water interaction (hydrogen bonds) becomes weaker in the vicinity of the surface, the larger ions (with respect to their crystallographic size) will be more attracted to the surface.

%%%%%%%%%%%%%%%%%%%%%%
\section{Concluding Remarks}
%%%%%%%%%%%%%%%%%%%%%%

Our work presents a general self-consistent theory for calculating free energies up to one-loop order
for ionic solutions with a surface proximal layer of finite width. In this layer, we consider a slowly varying ion-specific surface potential.
The loop expansion we use can be computed systematically to higher orders, and re-summing the loop terms
is actually equivalent to the cumulant expansion done in Refs.~\onlinecite{dean2003,dean2004}.

The calculation of the excess surface tension, $\Delta\gamma$, is based on the  free-energy difference
between a system with an air/water interface on one hand, and two semi-infinite systems (electrolyte and air)
with no interface on the other hand. The same calculation can be applied to other liquid/liquid interfaces such
as oil/water, simply by using $\varepsilon_a$ as the dielectric constant of the oil instead of air.
This calculation method is equivalent to the Gibbs absorbtion isotherm method~\cite{ll}, but it is mathematically
more accessible as it avoids the explicit calculation of the ionic densities.
It is also possible to use the grand potential instead of the  free-energy,
but the latter does not simplify the calculation, because one has to consider the one-loop correction to the fugacities (Appendix~C).

We have computed the linearized MFT electrostatic potential as well as the MFT and one-loop free energies,
utilizing the {\it secular determinant} method,
where we have extended the secular determinant method
of Ref.~\onlinecite{fdet} for the three boundaries as in our system.

The one-loop excess surface tension, $\Delta\gamma_1$, is calculated and the OS surface tension result was naturally recovered and extended.
In fact, we showed that the OS result is obtained by considering the
thermodynamic fluctuations of the electrostatic potential around its MFT solution,
while the volume fluctuations recovered the known DH correction to the MFT free energy.
Our surface-tension result is analytical and interpolates between several known limits:
the result of Ref.~\onlinecite{EPL} for $d \to 0$, the OS one for $\alpha_{\pm} \to 0$ or $d=0$,
and a Stern layer for $\alpha_{\pm} \to \infty$.
A wide variety of monovalent ions at the air/water and oil/water interfaces were fitted by our model,
all taken on a common and unified ground.

Within some approximations we obtain an analytical dependence of the excess surface tension
on the salt concentration.
The fits for $\Delta\gamma$ agree well with experiments and show clearly the
reversed Hofmeister series (F$^{-}>$\,Cl$^{-}>$\,Br$^{-}>$\,I$^{-}$) both at the air/water and oil/water interfaces.
The various fits reveal an even more extended (Hofmeister) series:
${\rm F}^{-} > {\rm IO_3}^{-}  > {\rm Cl}^{-} > {\rm BrO_3}^{-} > {\rm Br}^{-} > {\rm ClO_3}^{-} > {\rm NO_3}^{-} > {\rm I}^{-}$.
% > {\rm ClO_4}^{-}$.
In Ref.~\onlinecite{levin2010} a different series was obtained for anions:
${\rm IO_3}^{-} > {\rm F}^{-} > {\rm BrO_3}^{-} > {\rm Cl}^{-} > {\rm NO_3}^{-} > {\rm Br}^{-} > {\rm ClO_3}^{-} > {\rm I}^{-}$.
% > {\rm ClO_4}^{-}$
At present, it is not clear which of these two predictions is more accurate. Other experimental measurements, such as surface potential, might
shed light on this discrepancy. We intend to further investigate this issue in the future by calculating the surface potential to one-loop order.

In the weak-coupling linear regime, where our theory is valid,
fluctuations dominate over the MFT contribution,
and the cation and anion adhesivities are roughly equal, $\alpha_+ \simeq \alpha_-$.
This then leads to a small MFT contribution,
while the one-loop contribution has terms proportional to $\ln(\Lambda)$
and independent on $\alpha_{\pm}$.
These terms arise from the image charge interaction and, indeed,
give the OS result with an ionic-size correction (minimal distance of approach between ions).

Two important limitations of our theory are related to
the ion finite size,  and
to the linearization of the MFT  electrostatic equations.
For large adhesivities, where the ion density on the surface
is high and does not correspond to the
dilute solution limit,
our theory is expected to fail. Instead, other theories, such as the modified PB ones~\cite{MPB},
can be utilized as they take into account explicitly the ion finite size.
Considering, for example, the surface tension of acids such as HCl and HClO$_4$,
the fitted $\alpha$'s from our theory are found to be rather large (see Table~II) and
are not expected to be as reliable, for the reasons mentioned above.
Furthermore, when $|\alpha_+-\alpha_{-}|\simeq \kbt$,
the linear approximation fails and one should solve the full MFT electrostatic equation.

%%%%%%%%%%%%%%%% fig 6%%%%%%%%%%%%%%%%%%%%
%%%%%%%%%%%%%%%%%%%%%%%%%%%%%%%%%%%%%%%%%%%

%
Note that the image charge term (or other ion-interface interactions)
cannot be simply added in the Boltzmann weight factor (potential of mean force).
Since the PB equation is a MFT equation that follows from a certain free-energy minimization,
a consistent way to generalize it should be based on an augmented free-energy functional, which then gives a generalized PB equation.
In this way, double-counting of different electrostatic contributions is prevented,
and remedy a common ambiguity where the image charge or other ion-interface interactions
are added to the PB equation in an {\it ad hoc} fashion.

Our model provides a self-consistent way to calculate the fluctuations around the MFT free energy.
The image charge contribution naturally arises from the fluctuations hence, $\alpha_{\pm}$ originate only from solvent (short-range)
interactions and the problem of double counting is avoided.
An augmented free energy (``action''), Eq.~(\ref{m9}), is obtained and its minimization indeed leads to a modified PB equation.

The microscopic origin of the adhesivity is still not very well understood.
Nevertheless, several possibilities have been suggested recently.
For the special case of silica/water interface~\cite{Sivan2009,Sivan2013}
the orientation of water molecules in the vicinity of the interface
was proposed to lead to changes in the hydrogen bond strength at the interface.
This surface effect can be identified as a possible microscopic
source of $\alpha_{\pm}$, whose value is proportional to the difference in free energy
between a single ion solvated in the bulk as compared to its partially solvated state at the surface.
Another possible origin for the adhesivity is the cavitational energy~\cite{levin200y}.
The ion interferes with the water structure by breaking hydrogen bonds.
This costs energy, leading to an effective attraction of the ion to the interface.
Other effects such as the ion polarizability~\cite{levin200y} or dispersion forces~\cite{ninham2001}
might also play a role in the ion-surface interaction.
Since all proposed origins of the adhesivity are short range (their range is comparable to the ionic radius)
and our theory is a coarse-grained one, it is appropriate to average over their degrees of freedom and obtain $\alpha_{\pm}$,
whose magnitude is reasonably obtained from the fits to the experimental data.

Our model can be applicable for many systems which exhibit
a spatial region with ionic-specific slowly varying interactions.
Examples of such systems are, polymer-brushes densely
attached to a surface or polyelectrolyte gels.
Our model is ionic-specific and can be used to obtain
simple and analytical predictions for such complex systems.

Our model is applicable to many systems that exhibit a spatial region characterized by ionic-specific and slowly varying interactions. Two examples of
such systems are a densely layer of polymer brushes that are attached to a solid surface, or  a polyelectrolyte gel occupying a finite volume of solution. Our model
is ionic-specific and can be used to obtain simple and analytical predictions also for such complex inhomogeneous systems. 

%%%%%%%%%%%%%%%%%%%%%%%%%%%%%%%
%\section*{Acknowledgements}
%%%%%%%%%%%%%%%%%%%%%%%%%%%%%%%%
\acknowledgments
This work was supported in part by the Israel Science Foundation (ISF) under Grant No. 438/12 and the US-Israel Binational Science Foundation (BSF) under Grant No. 2012/060.
We thank M. Biesheuvel, Y. Ben-Yakar, A. Cohen,  H. Diamant, R. Netz, H. Orland and Y. Tsori
for numerous suggestions, and D. Frydel, Y. Levin and W. Kunz for  useful discussions.
One of us (RP) would like to thank Tel Aviv University for its hospitality during his stay there, and acknowledges the support of the ARRS through grant P1-0055.

%%%%%%%%%%%%%%%%%%%%%%%%%%%%%%%%%%%%%%%%%%%%%%%%%%%%
\appendix
%%%%%%%%%%%%%%%%%%%%%%%%%%%%%%%%%%%%%%%%%%%%%%%%%%%%
\section{Secular Determinant}
%%%%%%%%%%%%%%%%%%%%%%%%%%%%%%%%%%%%%%%%%%%%%%%%%%%%

The $ 4\times 4$ coefficient matrices $A$, $B$ and $C$ for the boundary condition equation, as discussed in Eq.~(\ref{ol5}), are:
\begin{eqnarray}
\label{sd1}
\nonumber& &A = \begin{pmatrix}
        -\varepsilon_a k \quad&\quad \!\!\!\!\!\!\!\!\varepsilon_w \quad&\quad \!\!0 \quad&\quad 0\,\,\\
        0                \quad&\quad \!\!\!\!\!\!\!\!0             \quad&\quad \!\!0 \quad&\quad 0\,\, \\
        0                \quad&\quad \!\!\!\!\!\!\!\!0             \quad&\quad \!\!0 \quad&\quad 0\,\, \\
        0                \quad&\quad \!\!\!\!\!\!\!\!0             \quad&\quad \!\!0 \quad&\quad 0\,\,
    \end{pmatrix} \, , \\ \nonumber\\
\nonumber& &B = \begin{pmatrix}
        \,\,\,\, 0 \quad&\quad 0 \quad&\quad 0            \quad&\quad 0\,\, \\
        \,\,\,\, 1 \quad&\quad 0 \quad&\quad \!\!\!\!\!-1 \quad&\quad 0\,\, \\
        \,\,\,\, 0 \quad&\quad 1 \quad&\quad 0            \quad&\quad \!\!\!\!\!-1\,\, \\
        \,\,\,\, 0 \quad&\quad 0 \quad&\quad 0            \quad&\quad 0\,\,
    \end{pmatrix} \, ,\\ \nonumber\\
& &C = \begin{pmatrix}
        \,\,\,\,0 \quad&\quad 0 \quad&\quad 0 \quad&\quad 0\,\, \\
        \,\,\,\,0 \quad&\quad 0 \quad&\quad 0 \quad&\quad 0\,\, \\
        \,\,\,\,0 \quad&\quad 0 \quad&\quad 0 \quad&\quad 0\,\, \\
        \,\,\,\,0 \quad&\quad 0 \quad&\quad 0 \quad&\quad 1\,\,
    \end{pmatrix}  \, .
\end{eqnarray}
In general, for a system with two boundaries, the secular determinant can be written as~\cite{fdet}
\begin{eqnarray}
\label{sd2}
\nonumber D_\nu({\bf k}) = \det\left[A + B \, \Gamma_\nu(a)\Gamma_\nu^{-1}(0) + C \, \Gamma_\nu(L)\Gamma_\nu^{-1}(0) \right] \, , \\
\end{eqnarray}
with $\Gamma_\nu(z)$ computed at the spatial point $z$ and is defined as another $4 \times 4$ matrix
\begin{eqnarray}
\label{sd3}
\nonumber& &\Gamma_\nu(z) = \begin{pmatrix}
        h^{(2)}_{\nu}   \quad&\quad  \!\!\!\!\!\!\!\!g^{(2)}_{\nu}    \quad&\quad \!\!\!\!\!\!\!\!0  \quad&\quad \!\!\!\!\!\!\!\!0\\
        \partial_z h^{(2)}_{\nu} \quad&\quad  \!\!\!\!\!\!\!\!\partial_z g^{(2)}_{\nu} \quad&\quad \!\!\!\!\!\!\!\!0  \quad&\quad \!\!\!\!\!\!\!\!0 \\
        0     \quad&\quad \!\!\!\!\!\!\!\!0    \quad&\quad \!\!\!\!\!\!\!\!h^{(3)}_{\nu}  \quad&\quad  \!\!\!\!\!\!\!\!g^{(3)}_{\nu} \\
        0     \quad&\quad \!\!\!\!\!\!\!\!0     \quad&\quad \!\!\!\!\!\!\!\!\partial_z h^{(3)}_{\nu}  \quad&\quad  \!\!\!\!\!\!\!\!\partial_z g^{(3)}_{\nu}
    \end{pmatrix}   \, .
\end{eqnarray}
In the equation above we have used the notation $h^{(i)}_{\nu}(z)$ and $g^{(i)}_{\nu}(z)$ with $i=2,3$ for the
two independent solutions of the eigenvalue equation, Eq.~(\ref{ol3}), in the proximal ($0 \leq z \leq d$)
and distal ($z>d$) regions, respectively.
It is then convenient to choose these independent solutions such that $\Gamma(0) = 1$.
The secular determinant expression, Eq.~(\ref{sd2}), can be then simplified  as is used in Eq.~(\ref{ol9}).

%%%%%%%%%%%%%%%%%%%%%%%%%%%%%%%%%%%%%%%%%%%%%%%%%%%%
\section{One-loop Surface Tension to Order $O(1/\Lambda)$}
%%%%%%%%%%%%%%%%%%%%%%%%%%%%%%%%%%%%%%%%%%%%%%%%%%%%

The analytical solution to the integral in Eq.~(\ref{so2a}) in the limit $\Lambda\rightarrow\infty$,
can be computed be neglecting terms of order $O(1/\Lambda)$,
while retaining constant terms (with respect to $\Lambda$):
\begin{eqnarray}
\label{b1}
\nonumber&\Delta\gamma_1 &\simeq -\frac{\kbt}{8\pi}\left(\frac{\varepsilon_w-\varepsilon_a}{\varepsilon_w+\varepsilon_a}\right) \frac{\kd^2}{4} \Bigg[ 2\ln (\kd \Lambda^{-1}) - 1 \Bigg.\\
\nonumber & &\Bigg.+ \frac{4\varepsilon_w}{(\varepsilon_w-\varepsilon_a)^2}\left(\varepsilon_a\ln 2 - \varepsilon_w\ln \frac{\varepsilon_w+\varepsilon_a}{\varepsilon_w+\omega \kd^{-1}}\right) \Bigg] \\
\nonumber & & + \frac{\kbt}{8\pi}\frac{\omega \kd}{\varepsilon_w+\varepsilon_a} \left[- \frac{\varepsilon_w}{\varepsilon_w-\varepsilon_a} - \frac{\omega}{2\kd\left(\varepsilon_w+\varepsilon_a\right)} \right.\\
\nonumber & &\left.+ \frac{\omega}{\kd \left(\varepsilon_w-\varepsilon_a\right)}\left(\frac{2\varepsilon_a}{\varepsilon_w-\varepsilon_a}\ln 2 - \ln \frac{\varepsilon_w+\varepsilon_a}{\varepsilon_w + \omega\kd^{-1}} \right) \right.\\
\nonumber & &\left. + \frac{\omega}{\kd\left(\varepsilon_w+\varepsilon_a\right)}\ln (\kd \Lambda^{-1}) \right.\\
\nonumber & &\left.+ \frac{4\varepsilon_w\varepsilon_a\left[\kd^2\left(\varepsilon_w^2-\varepsilon_a^2\right) - \omega^2\right]^{\half}}{\kd\left(\varepsilon_w-\varepsilon_a\right)^2\left(\varepsilon_w+\varepsilon_a\right)} \right.\times\\
& &\qquad\times \left.\arctan (\frac{\sqrt{\kd^2\left(\varepsilon_w^2-\varepsilon_a^2\right)-\omega^2}}
{\kd\left(\varepsilon_w+\varepsilon_a\right) +\omega}     \right] \, ,
\end{eqnarray}
with $\omega = \varepsilon_w d\left(\xi^2-\kd^2\right)$.
The first term above (proportional to $\kd^2$) represents a correction to the OS result.
In the case where $\alpha_{\pm} = 0$, we recover exactly the result obtained in Ref.~\onlinecite{podgornik1988} (apart from an inconsequential typo).
%The only difference is the second term in the first square brackets which is probably a typo in Ref.~\cite{podgornik1988}.
The second term (proportional to $\omega\kd$) is the ion-specific term, with leading behavior as in Eq.~(\ref{so3}).

%%%%%%%%%%%%%%%%%%%%%%%%%%%%%%%%%%%%%%%%%%%%%%%%%%%%
\section{Fugacities Second-order Correction}
%%%%%%%%%%%%%%%%%%%%%%%%%%%%%%%%%%%%%%%%%%%%%%%%%%%%
The grand-potential is written in terms of the fugacities, $\lambda_i$, with $i=\pm$, for the anions and cations, respectively.
The fugacities are related to the bulk densities by:
\begin{eqnarray}
\label{e19}
n^{(i)}_b = -\lambda_i\frac{\beta}{V}\frac{\partial\Omega}{\partial \lambda_i} \, .
\end{eqnarray}
It is clear from the above equation that if one modifies the grand potential with quadratic fluctuations,
the fugacities will be modified as well.
Writing the grand potential to one-loop order yields
\begin{eqnarray}
\label{e20}
\Omega \simeq \Omega_0(\lambda_i) + {\cal C}\Omega_1(\lambda_i) \, ,
\end{eqnarray}
where we introduce the parameter ${\cal C}$ to keep track of the expansion terms.
After finishing the calculation this parameter will be set to unity.
Substituting Eq.~(\ref{e20}) into Eq.~(\ref{e19}) gives, $\lambda_{i,1}$, the one-loop correction to the fugacities:
\begin{eqnarray}
\label{e21}
\nonumber\lambda_i &\simeq& \lambda_{i,0} + {\cal C}\lambda_{i,1} \\
&=& -\left[ \frac{V n^{(i)}_b}{\beta\partial\Omega_0/\partial \lambda_i} + {\cal C}\lambda_{i,0}\Bigg.
\frac{\partial\Omega_1/\partial \lambda_i}{\partial\Omega_0/\partial \lambda_i}\right]_{\lambda_i=\lambda_{i,0}} \!\!\!\! ,
\end{eqnarray}
where $\lambda_{i,0}$ is the zeroth-order fugacity.
We now expand the grand-potential correction, $\Omega_1(\lambda_i)$, around $\lambda_{i,0}$
and use Eq.~(\ref{e21}) for $\lambda_{i,1}$ to obtain
\begin{eqnarray}
\label{e23}
\nonumber&\Omega &\simeq \Omega_0(\lambda_{i,0}) \\
& &+ \, {\cal C}\left[\Omega_1(\lambda_{i,0}) - \sum_{i=\pm} \lambda_{i,0} \Bigg.\frac{\partial
\Omega_1(\lambda_i)}{\partial \lambda_i} \right]_{\lambda_i=\lambda_{i,0}} \!\!\!\! .
\end{eqnarray}
The electro-neutrality condition $\sum_i \lambda_iq_i=0$, for symmetric electrolytes, imposes
$\lambda_+=\lambda_-\equiv \lambda$.
Using Eq.~(\ref{e19}) and the definition of the fugacities, Eq.~(\ref{m7}),
we obtain the intrinsic chemical potential:
\begin{eqnarray}
\label{e24}
\mu_i(\vecr) \simeq -\half e^2u({\bf r,r}) + \kbt\left[\ln \left(\lambda_{i,0} a^3\right) +
{\cal C}\frac{\lambda_{i,1}}{\lambda_{i,0}}\right] \, . \qquad
\end{eqnarray}
In the above equation we have also expanded the fugacities around $\lambda_{i,0}$.
We now calculate the free energy from Eqs.~(\ref{f2}) and (\ref{e23}),
with the chemical potential obtained in Eq.~(\ref{e24}),
\begin{eqnarray}
\label{e25a}
& & {\cal F} \simeq {\cal F}_0 + {\cal C}{\cal F}_1 = \Omega_0(\lambda_{i,0}) -\frac{e^2}{2} n_b^{(i)} V u_{b}({\bf r,r}) \\
\nonumber & &+ \, V\kbt\sum_{i=\pm}  n_b^{(i)}\ln \left(\lambda_{i,0}a^3\right) \, + \, {\cal C}\Bigg[\Omega_1(\lambda_{i,0}) + \, \Bigg. \\
\nonumber& & \left.+ \, \sum_{i=\pm} \lambda_{i,1} \left(\frac{\partial\Omega_0(\lambda_i)}{\partial \lambda_i}\right)_{\lambda_i=\lambda_{i,0}}
\!\!\!\!\!\!\!\! + \, V\kbt\sum_{i=\pm} n_b^{(i)}\frac{\lambda_{i,1}}{\lambda_{i,0}} \right] \, .
\end{eqnarray}
The second and third terms in ${\cal F}_1$ (the square brackets) cancel each other exactly
(from the definition of $\lambda_0$, Eq.~(\ref{e21})).
Thus, the free-energy to first order yields,
\begin{eqnarray}
\label{e26}
\nonumber& {\cal F} &\simeq \Omega_0(\lambda_{i,0}) -\half q^2 n^{(i)}_b V u_{b}({\bf r,r}) \\
& &+ V\kbt\sum_{i=\pm}  n^{(i)}_b\ln \left(\lambda_{i,0} a^3\right) + \Omega_1(\lambda_{i,0}) \, .
\end{eqnarray}
This result gives rise to a useful simplification,
where to first order in a loop expansion of the free-energy,
the fugacities can be taken as $\lambda_{i,0}$,
and are equal to the bulk densities.
Note that this is not the case for the grand potential, $\Omega$.
For the latter, the fugacities must be calculated consistently to one-loop order as shown in Eq.~(\ref{e21}).

%%%%%%%%%%%%%%%%%%%%%%%%%%%

%%%%%%%%%%%%%%%%%%%%%

%%%%%%%%%%%%%%

\begin{thebibliography}{99}
\bibitem{foot1}
Taking the $\alpha_{\pm}\to\infty$ limit means that ions are excluded from the proximal $[0,d]$ region.
It is clear that one only needs to take $q\to k$ and then the secular determinant assumes a form corresponding to the Stern layer,
%
\begin{eqnarray}
\label{ol17}
\nonumber& D(k) \simeq& \e^{pL} \e^{(k-p)d}\left(\varepsilon_w + \varepsilon_a \right) \\
\nonumber& &\times \Bigg[ k\Big( 1 - \Delta \e^{-2kd} \Big) + p\Big( 1 + \Delta \e^{-2kd} \Big) \Bigg]  \, ,
\end{eqnarray}
%
where $\Delta \equiv \Delta(k,k) = \left( \varepsilon_w - \varepsilon_a \right)/\left( \varepsilon_w + \varepsilon_a \right)$.
\bibitem{foot3}
An alternative derivation is presented in Ref.~\onlinecite{EPL} and can be obtained from our
Eq.~(\ref{ol18}) by identifying $\omega = \varepsilon_w d\left(\xi^2-\kd^2\right)$.
\end{thebibliography}
\end{document}